\documentclass[onecolumn]{aa} 

\usepackage{graphicx}
\usepackage{txfonts}
\usepackage{natbib}
\begin{document}
   \title{The structure of the nuclear stellar cluster of the Milky Way}

   \subtitle{}

   \author{R. Sch\"odel\inst{1}
           \and
          A. Eckart\inst{1}
          \and
          T. Alexander\inst{4,5}
          \and
          D. Merritt\inst{6}
          \and
          R. Genzel\inst{2,3}
          \and
          A. Sternberg\inst{7}
          \and
          L. Meyer\inst{1}
          \and
          F. Kul\inst{1}
          \and
          J. Moultaka\inst{8}
          \and
          T. Ott\inst{2}
          \and
          C. Straubmeier\inst{1}
          }

   \offprints{R. Sch\"odel}

   \institute{I.Physikalisches Institut, Universit\"at zu K\"oln,
              Z\"ulpicher Str.77, 50937 K\"oln, Germany\\ 
              \email{rainer,eckart,leo,kul,cstraubm@ph1.uni-koeln.de}
         \and
            Max-Planck-Institut f\"ur extraterrestrische Physik,
  Giessenbachstra{\ss}e, 85748 Garching, Germany\\ 
  \email{genzel,ott@mpe.mpg.de}
\and 
Also: Department of Physics, University of
  California, Berkeley, CA 94720, USA
\and 
Faculty of Physics, Weizmann Institute of Science,
              76100 Rehovot, Israel\\
  \email{tal.alexander@weizmann.ac.il}
\and
Incumbent of the William Z.\ \& Eda Bess Novick career development chair
\and
Department of Physics,  Rochester Institute of Technology, 
54 Lomb Memorial Drive, Rochester, NY 14623-5604\\
\email{merritt@astro.rit.edu}
\and
School of Physics and Astronomy and the Wise Observatory, The Beverly
              and Raymond Sackler Faculty of Exact Sciences, Tel Aviv
              University, Tel Aviv 69978, Israel\\
\email{amiel@wise.tau.ac.il}
\and
Laboratoire d'Astrophysique de Toulouse, UMR 5572, Observatoire
              Midi-Pyr\'en\'ees, 14 avenue Edouard Belin, 31400
              Toulouse, France\\
\email{jihane.moultaka@ast.obs-mip.fr}
 }

   \date{}

  \abstract 
  {}  
  {The centre of the Milky Way is the nearest nucleus of a galaxy and
  offers a unique possibility to study the structure and dynamics of a dense
  stellar cluster around a super-massive black hole.}
   {We present high-resolution seeing limited and AO  NIR imaging 
observations of the stellar cluster within about one parsec of Sgr~A*, the
massive black hole at the centre of the Milky Way. Stellar number
counts and the diffuse background light density
 were extracted from these observations in order to examine the
structure of the nuclear stellar cluster.  A detailed map of the variation of
  interstellar extinction in the central $\sim0.5$\,pc of the Milky Way is presented and used to correct the stellar number counts and diffuse light
  density.}
   {Our findings are as follows: (a) A broken-power law
provides an excellent fit to the overall structure of the GC
nuclear cluster. The power-law slope of the cusp
is $\Gamma=0.19\pm0.05$, the break radius is $R_{\rm break} = 6.0''
\pm 1.0''$ or $0.22\pm0.04$\,pc, and the cluster density decreases
with a power-law index of $\Gamma=0.75\pm0.1$ outside of $R_{\rm
break}$.  (b) Using the best velocity dispersion measurements from the
literature, we derive higher mass estimates for the central parsec than assumed
until now. The inferred density of the cluster at the break radius is $2.8\pm1.3\times 10^{6} {\rm M_{\odot}\,pc^{-3}}$. This high density agrees well with the small extent and
flat slope of the cusp. Possibly, the mass of the  stars makes
  up only about 50\% of the total cluster mass.  (c) Possible
indications of mass segregation in the cusp are found (d) The cluster
appears not entirely homogeneous. Several density clumps are detected
that are concentrated at projected distances of $R=3''$ and $R=7''$
from Sgr~A*.(e) There appears to exist an under-density of horizontal
branch/red clump stars near $R=5''$, or an over-density of stars of
similar brightness at $R=3''$ and $R=7''$.  (f) The extinction map in
combination with cometary-like features in an L'-band image may
provide support for the assumption of an outflow from Sgr~A*. }
{}


   \maketitle

%

\section{Introduction}

The centre of the Milky Way is the closest galactic nucleus, located
at a distance of about 8\,kpc
\citep{Reid1993ARA&A,Eisenhauer2003ApJ}. It represents the only case
where it is possible to observe and analyse the interplay between a
central massive black hole
\citep[e.g.,][]{Eckart1996Natur,Genzel2000MNRAS,Ghez2000Natur,Schoedel2002Natur,Ghez2003ApJ}
and the gas and stars in its environment directly through observation
of individual stellar sources and gas features. A particular point of
interest is the density structure of the nuclear star cluster. From
theoretical considerations one expects to observe a stellar cusp
around the black hole.

The way in which a central black hole alters the structure of a
surrounding cluster has been examined theoretically starting from two
fundamentally different assumptions: On the one hand, there are
adiabatic models, in which the growth rate of the black hole is long
compared to the dynamical time scale of the cluster, but short with
respect to the two-body relaxation time of the cluster. Under these
assumptions, a $r^{-3/2}$-cusp will develop in a spherically
symmetric, non-rotating system that is initially isothermal
\citep{Young1980ApJ}. In the case of rotating systems and
non-isothermal clusters, cusps as steep as $r^{-5/2}$ may develop
\citep{LeeGoodman1989ApJ,Quinlan1995ApJ}. On the other hand, the
system may be older than the two-body relaxation
time. \citet{Bahcall1976ApJ} found that in this case, the final
configuration will be a $r^{-7/4}$-cusp for a single-mass stellar
population, independent of the initial
conditions. \citet{BahcallWolf1977ApJ} extended their work to a system
containing stars of two different masses and found that the
steady-state distribution near the black hole approaches a power-law
with index:
\begin{equation}
\gamma=\frac{m_{1}}{4m_{2}}+\frac{3}{2}, m_{1}\leq m_{2}.
\end{equation}
Therefore, in a stellar cluster with a range of stellar masses, one
would expect a slope with $\gamma$ in the range $3/2$ to $7/4$ (From
here on, we will use $\gamma$ as notation for the power-law index of
the space density and $\Gamma$ for the power-law index of the
projected stellar density, the quantity that is actually observed).
\citet{Murphy1991ApJ} simulated the evolution of multi-mass clusters
with a Fokker-Planck code. They also found that the slope of the final
cusp will lie in the range $-7/4$ to $-1/2$ \citep[see
also][]{LightmanShapiro1977ApJ}. The latter value represents the
case of high-density nuclei, where stellar collisions become important
and lead to a flattening of the cusp.

Hence, a stellar cusp is expected to be the signature of a central
black hole embedded in a stellar cluster. The properties of the cusp,
particularly the index of the power-law that describes the stellar
density, should reflect the formation history of the system.

Unfortunately, only the structure of the light distribution can be
examined in external galaxies because of the lack of sensitivity and
resolution of the current generation of instruments. Surface
brightness measurements, however, may be dominated by the light from a
few extremely bright stars and may therefore be only of limited use
when examining the structure of nuclear clusters. Due to its
proximity, the GC nuclear cluster can be resolved into individual
sources and may represent therefore an ideal test case for cusp
theories.

In one of the early attempts to determine the density
structure of the stellar cluster around the GC,
\citet{Catchpole1990MNRAS} counted giant stars in a
$1^{\circ}\times2^{\circ}$ region centred on the GC. They found that
the increase of the stellar volume density toward the centre was about
$\propto r^{-2}$, as in an isothermal cluster, where $r$ is the
distance to the GC.  \citet{Haller1996ApJ} determined the surface
brightness profile at distances between $15''$ and $200''$ from Sgr~A*
and found a dependence $\propto R^{-0.8}$, where $R$ denotes the
projected distance from the centre of the cluster \citep[see
also][]{BecklinNeugebauer1968ApJ}. This corresponds to a density
distribution $\propto r^{-1.8}$. 

\citet{Eckart1993ApJ} and \citet{Genzel1996ApJ} presented number
density counts from high-resolution near-infrared (NIR) speckle
imaging observations at the diffraction limit of a 4\,m-class
telescope in a region from about $1''$ to $20''$. They found that the
stellar density appears to be described well by an isothermal cluster
($\rho\propto r^{-2}$). In the inner few arc-seconds the surface
density distribution was found to flatten. The core radius of the
cluster was determined to $\sim$$0.15-0.4$\,pc
\citep{Eckart1993ApJ,Genzel1996ApJ}. Some indication for a cusp, i.e.\
a central density excess above the flat core of an isothermal cluster
was found in follow-up work
\citep[e.g.,][]{Eckart1995ApJ,Alexander1999ApJ}, but the evidence
remained somewhat inconclusive. At this point, all attempts to
determine the structure of the GC nuclear cluster were limited to
stars brighter than $\mathrm{mag_{K}}=15$ and to resolutions of a few
$0.1"$ (with the exception of the speckle imaging).

This situation changed with the commissioning of the adaptive optics
(AO) module NAOS and the NIR camera CONICA (the combination of which
is abbreviated NACO) at the ESO 8\,m-class VLT unit telescope~4 on
Paranal, Chile. The Paranal mountain is an ideal site for observations
of the GC, which passes close to zenith at this
location. Additionally, the unique NIR wavefront sensor of NAOS allows
to lock the AO on the NIR bright supergiant IRS~7 that is located just
$\sim$$6''$ from Sgr~A*.  With this instrument, imaging observations
can be obtained with a resolution of the order 50~milli-arc-seconds
(mas) that are about 2-3 magnitudes deeper than previous speckle
imaging observations.  \citet{Genzel2003ApJ} used the first NACO H-
and K-band imaging data to present a new analysis of the stellar
number density in the GC. They combined high-resolution number counts
from NACO data in the region from $0.1''$ to $10''$ projected radius
from Sgr~A* with lower resolution number counts from speckle imaging
observations of fields at distances between $10''$ and $100''$. The
combined surface density counts vs.~distance from Sgr~A* showed that
the counts could be described reasonably well by an isothermal cluster
at distances greater than $5-10''$. They could show, for the first
time, clear evidence for a density excess above a flat core within
$\sim$2$''$ of Sgr~A*. By fitting a broken power-law to the number
density counts a power law index of $\gamma=-1.4\pm0.1$ was determined
for the stellar space density distribution in the cusp. However, they
also find that a large number of the stars with $\mathrm{mag_{K}} \leq
15$ inside a projected radius of $10''$ appear to be of early
type. The presence of these massive, young stars indicates that the
visible part of the GC cluster may not be dominated by a stellar
population that is old and had time to settle into a relaxed state as
is assumed in many cusp formation theories. It seems that at least
among the stars brighter than $\mathrm{mag_{K}} = 15$ a significant
fraction belongs to the discs of young, massive stars discovered in
the inner few arc seconds around Sgr~A* \citep[e.g.,
see][]{Paumard2006ApJ}. \citet{Scoville2003ApJ} find a flatter
brightness profile than \citet{Genzel2003ApJ} from analysing the
surface brightness in HST images. Clearly, the question on which is
the best technique to measure the structure of the GC stellar cluster
-- surface brightness measurements or number counts -- must be
addressed.

The present work builds up on and expands the work of
\citet{Genzel2003ApJ}. The differences/new techniques are in
particular: a) Use of a new, homogeneous data set. All high-resolution
imaging data in this work are from AO observations with
NACO/VLT\footnote{Based on observations collected at the European
Southern Observatory, Chile, programs 073.B-0745 and 075.C-0138.} in
the K-band. They are considerably less affected by saturation than the
data in our previous work and reach out to a projected radius of about
$20''$, well beyond the extent of the cusp \citep[the AO data
in][reached only out to $10''$]{Genzel2003ApJ}. We do not need to
combine star counts from different sources. b) High-quality seeing
limited images are combined with the AO imaging data. They provide
independent estimates of star counts and of the background light
density, which cannot be determined reasonable in the AO images.  c)
PSF fitting with a spatially variable PSF is applied to optimise the
detection of stellar sources. d) The diffuse background light is
analysed in order to determine the structure of the faint, unresolved
stellar component of the cluster. e) Various methods are used to
extract the shape of the underlying cluster from the star
counts. Binning of the star counts is avoided. Non-parametric methods
are applied to derive the surface and space density of the cluster.
We address f) the problem of surface brightness vs.\ star counts. g)
The star counts and background light density analyses are corrected
for extinction. h) Besides focusing only on the radially averaged
large scale structure of the cluster, the two-dimensional structure of
the cluster is examined as well. Significant inhomogeneities of the
cluster are detected at several locations. i) The surface density of
stars is derived for different stellar magnitude ranges, which makes
it possible to search for effects of mass segregation and to probe the
old, most probably dynamically relaxed stellar population that is
represented by horizontal branch/red clump stars.

Throughout the paper we will adopt a GC distance of $7.6$\,kpc
\citep{Eisenhauer2005ApJ}. At this distance, one arc second
corresponds to $0.037$\,pc.

\section{Observations and data reduction}

\subsection{ISAAC imaging \label{isaac}}

The GC was observed with the NIR camera and spectrometer ISAAC at the
ESO VLT unit telescope~4 on Paranal on 2 July 1999 and on 27 July
2005. Filters used were a narrow band filter of 0.02\,$\mu$m width
centred at 2.09\,$\mu$m and the J-band filter in 1999, the H-band
filter in 2005. Details of the observations are given in
Tab.~\ref{Tab:ISAAC}.

\begin{figure*}[!p]
\centering
\includegraphics[width=\textwidth]{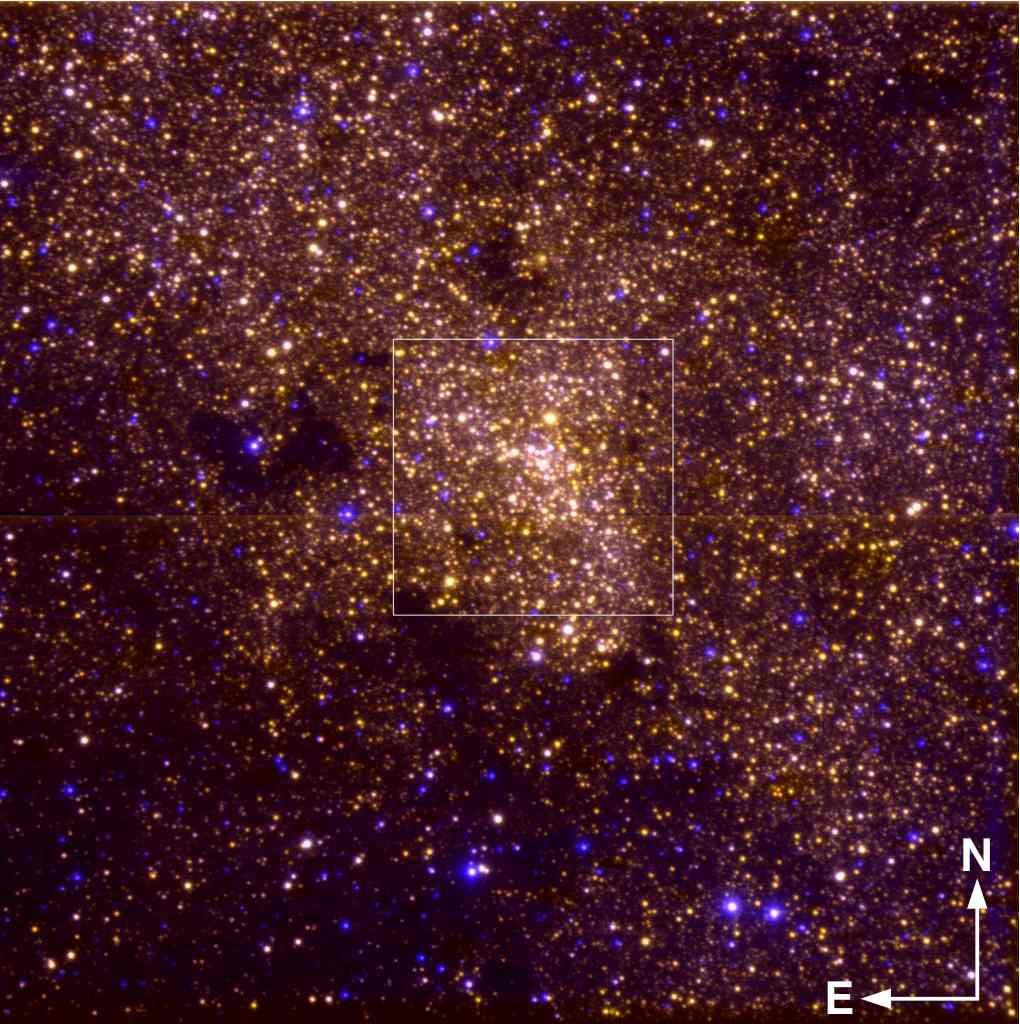}
\caption{Colour image composed of ISAAC imaging observations at
  $2.09$\,$\mu$m and in the J-band. The field-of-view is
  $150''\times150''$. The field of about $40''\times40''$ that was
  observed with AO observations is marked by a square. The galactic
  plane runs approximately southwest-northeast across the image. Blue
  sources are foreground stars. The patchy and highly variable
  extinction is evident, as well as the minimum of the extinction on
  the central cluster.
\label{Fig:ISAAC}
}
\end {figure*}

The data were sky subtracted and corrected for bad
pixels. Figure~\ref{Fig:ISAAC} shows a colour image composed of the
ISAAC J and K observations (which have both a similar PSF FWHM, while
the FWHM due to seeing in the H-band observations was about twice as
large). The colour image illustrates the strong, patchy and highly
variable extinction in the GC. It shows also that there is a general
minimum of the extinction on the central cluster and in a band running
north-east to south-west across it, as described by
\citet{Scoville2003ApJ}.

\subsection{AO imaging data}

The field-of-view (FOV) of the AO observations is marked with a white
square in Fig.~\ref{Fig:ISAAC}.

Various sets of data were chosen for the analysis of the stellar
number density and the interstellar extinction in the central
parsec. The data were selected according to criteria such as seeing,
size of the isoplanatic angle, quality of AO correction, coverage of a
large FOV, and the central filter wavelength. A summary of all 
data can be found in Tab.~\ref{Tab:Data}.

When imaging the field centred on Sgr~A*, the NIR wavefront sensor of
NAOS was locked on the bright supergiant IRS~7, located about 6$''$
north of Sgr~A*. Mosaic imaging was applied by using a rectangular
dither pattern in order to enlarge the FOV.  Several dithered
exposures of a dark cloud near the GC ($713''$ west, $400''$ north of
Sgr~A*), a region largely devoid of stars, were interspersed with all
observations in order to sample the sky background.

\begin{figure*}[!p]
\centering
\includegraphics[angle=270,width=\textwidth]{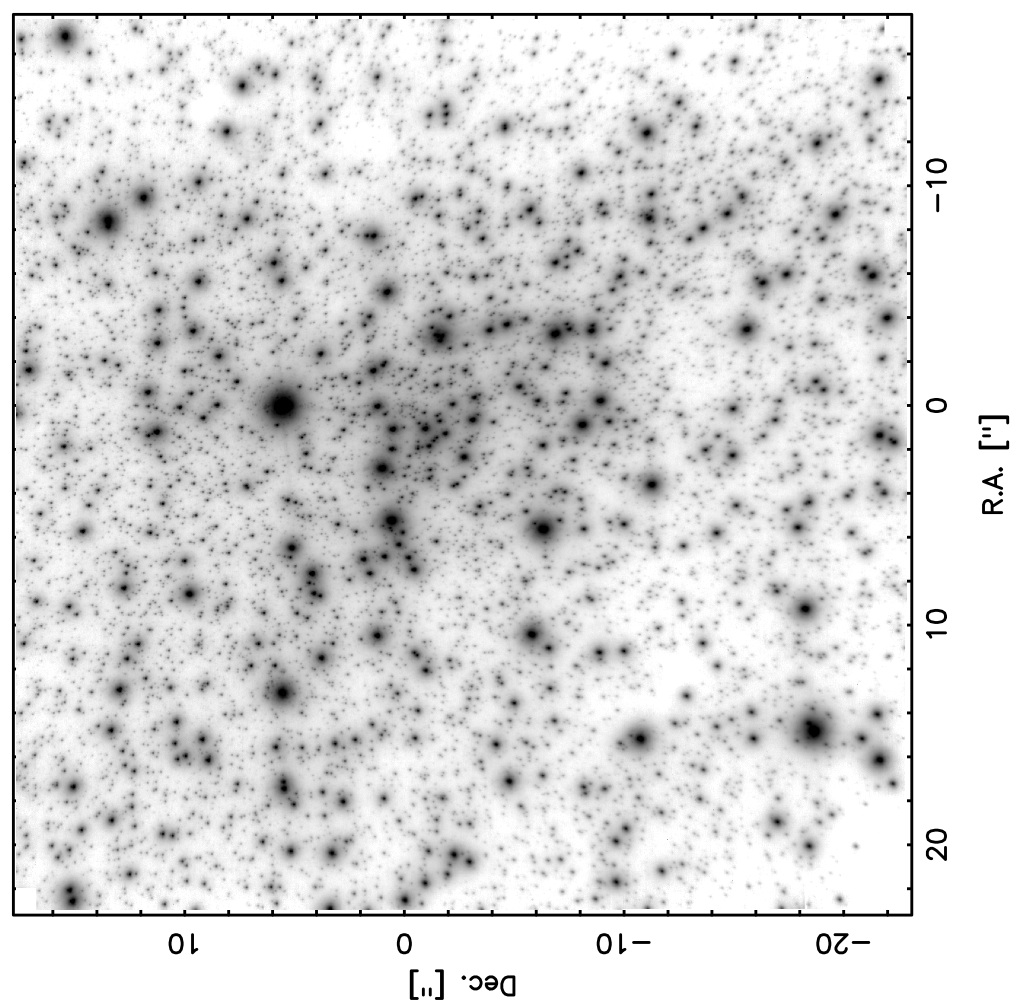}
\caption{Final mosaic of the $2.27 + 2.30$~$\mu$m IB filter AO imaging
data with the $0.027''$ pixel scale, centred on Sgr~A*.  The offsets
from Sgr~A* in right ascension and declination that are labelled on
the coordinate axes serve as an orientation, but do not represent high
accuracy astrometry because there may be a slight ($<1^{\circ}$)
remnant rotation present in the images. North is up and east is to the
left.
\label{Fig:NACO}
}
\end {figure*}

In all cases, the images were sky subtracted, flat-fielded, and
corrected for dead/bad pixels during data reduction. After shifting
the dithered images to a common position, they were median combined to
obtain a final mosaic. In Fig.~\ref{Fig:NACO} we show the sum image of
the $2.27$\,$\mu$m and $2.30$\,$\mu$m intermediate band (IB) filter
data of the central field.

\section{Structure of the cluster}

\subsection{Seeing limited observations}

The IDL program package \emph{StarFinder} \citep{Diolaiti2000A&AS} was
used for source detection, photometry, and background
estimation. Photometric calibration was based on stellar magnitudes
given by \citet[][K-band: IRS~16NE, IRS~33N, IRS~33SW, ID~62,
ID~96]{Ott1999ApJ} and \citet[][H-band: IRS~16NE, IRS~33N,
IRS~33SW]{Blum1996ApJ}. Selection criteria for these stars were that
the calibration stars should be isolated and bright, but not variable.

\subsubsection{Extinction \label{sec:ISAAC_ext}}

When examining the structure of the central stellar cluster, it is
desirable to take interstellar extinction into account. For the ISAAC
images, a simple approach was taken by assuming that all stars in the
FOV have the same intrinsic colour of $H-K=0.15$.

How well justified is the assumption of a single intrinsic stellar
colour?  The intrinsic $H-K$ colours of normal stars span a small
range in magnitude. More than 94\% of the stars in our seeing-limited
sample have $\rm mag_{K} > 12$. They are therefore either giants or
main sequence stars.  Main sequence stars in the considered magnitude
range must be of class O or B. Such stars are only observed in
significant numbers in the innermost arcsecond near Sgr~A*
\citep[][]{Eisenhauer2005ApJ,Paumard2006ApJ}. Assuming that all stars
are early K giants will therefore introduce an uncertainty $<0.1\,{\rm
mag}$ in the measured colour for more than 90\% of the stars.

Median stellar colours for the ISAAC images were obtained from the
measured individual stellar colours by taking the median of the 25
nearest stars at each position in the image. In this way, the effect
of any possible individual erroneous measurements was minimised.
Extinction values were then derived from the colours by applying the
Draine extinction law, $A_{\lambda}\propto\lambda^{-1.75}$
\citep{Draine1989book}. The consequence of this procedure is that the
resolution of the obtained extinction map ranges between $\sim$2.5$''$
in the densest parts of the cluster, near Sgr~A*, to $\sim$10$''$ near
the edge of the image. The spatial resolution decreases down to
$\sim$$15''$ in the southwestern corner, where few stars were detected
due to the presence of a large dark cloud. The statistical uncertainty
of the derived extinction values is $\leq 0.2$\,mag in more than 90\%
of the map and never higher than $0.3$\,mag. The systematic error of
the extinction map epends on the uncertainty of the absolute
calibration. An error in the absolute calibration will have no
influence on the determined {\it shape} of the stellar cluster because
it will only shift the counts/light densities by a fixed factor.  The
absolute calibration was chosen such that $\rm A_{K} \approx2.8$ near
Sgr~A* \citep{Eisenhauer2005ApJ}.

\begin{figure*}[!p]
\centering
\includegraphics[width=\textwidth]{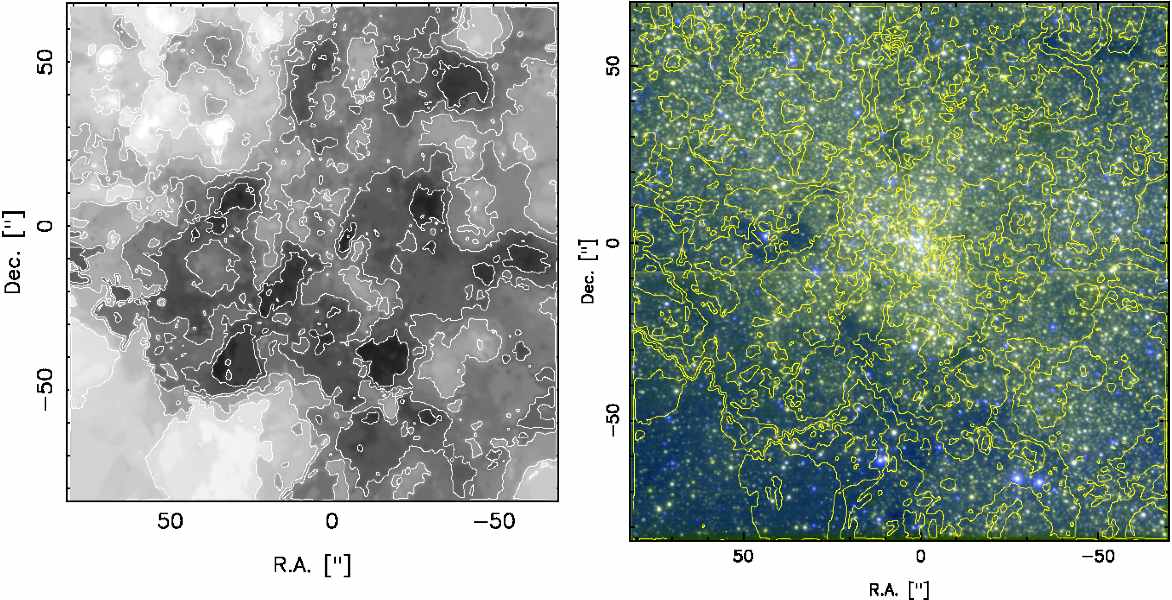}
\caption{Left: Extinction map derived from H-band and $2.09$\,$\mu$m
  images of the central stellar cluster with ISAAC/VLT. The details of
  the method are described in the text. Contours are plotted from $\rm
  A_{K}=1.8-3.6$ in steps of $\rm \Delta A_{K} = 0.3$. Darker shades
  correspond to higher extinction. Right: The extinction contours from
  the left panel are plotted over a combined J and K false colour
  image. Good agreement can be found between high extinction as seen
  by eye (darkness of regions, number of stars visible in regions) and
  the extinction contours. A stark discrepancy exists only in the
  south-western corner of the image. Here, extinction is so strong,
  that preferentially foreground stars contribute to the measurements,
  i.e.\ the extinction is underestimated.
\label{Fig:ISAAC_extinction}
}
\end {figure*}

The map of the interstellar extinction derived in this way is shown in
the left panel of Fig.~\ref{Fig:ISAAC_extinction}. The right panel
shows the extinction contours overlaid onto the combined J+K false
colour image.  It can be seen that darker regions and regions of low
stellar density correlate well with regions of increased
extinction. This test by eye confirms the general validity of the
approach. However, each method to measure extinction has its
drawbacks. A general limitation, certainly applicable in this case, is
that one sees only the projection of the target on the sky and lacks
three-dimensional information. In regions where the extinction is so
high that almost only foreground stars are detected, e.g.\ the
southwestern region, extinction will therefore be underestimated
significantly.  However, we do not include the problematic regions in
our analysis, which is limited to projected distances $R\leq60''$ from
Sgr~A*.

The extinction map presented here compares favourably with the map
presented by \citet{Scoville2003ApJ}. Very similar features can be
found, such as a general minimum of the extinction in a strip
southwest-northeast across the center and increased extinction in the
area of the circum-nuclear disc. Even details, like the linear
southeast-northwest running feature at about ($10''$,$-15''$), with a
lower extinction area to the northeast, agree well.

From the extinction map, we derived maps for the correction of stellar
counts and background light density. While this correction is
straightforward for the light density, assumptions on the luminosity
function must be made in order to correct the star counts.  Hence, the
counts at a given location were multiplied with a correction factor
that was derived by assuming a power law index of the stellar
luminosity function of $\beta=0.23$, derived from the KLF (see
section~\ref{sec:KLF} and Fig.~\ref{Fig:LF}).

\subsubsection{Star counts beyond 0.5\,pc \label{sec:ISAAC_counts}}

The analysis of the star counts was done on the 2.09\,$\mu$m image
because of the low extinction in this band and because of the high
resolution.  Seeing was excellent ($\sim$$0.4''$) and stable during
the 2.09\,$\mu$m observations. Contrary to observations with adaptive
optics, seeing limited observations do not suffer from PSF variations
due to the isoplanatic angle. Therefore it is possible to extract a
PSF that is valid for the entire FOV. More than $20,000$ stars were
detected in the FOV of the ISAAC 2.09\,$\mu$m observations. We limited
our analysis to a region within $\sim$$60''$ of Sgr~A* in order to be
able to extract counts in full rings around the centre (the FOV is not
centred on Sgr~A*, see Fig.~\ref{Fig:ISAAC}) and in order to avoid the
region of extremely high extinction to the southwest of Sgr~A*, for
which the extinction could not be accurately measured (see
Fig.~\ref{Fig:ISAAC_extinction}). The completeness of the star counts
was determined via the technique of introducing artificial stars into
the image and trying to re-detect them. The image was probed on a
dense grid of $\sim$$0.74''\times0.74''$. In order to avoid
artificially increasing the crowding, the procedure of introducing and
recovering artificial stars was repeated various times with a
sufficiently large grid (meaning the distance between artificial point
sources was greater than several times the PSF FWHM) that was shifted
with each try. Table~\ref{Tab:ISAAC_completeness} lists the
completeness levels of the ISAAC 2.09\,$\mu$m image for the entire
inner $60''$, for the inner $15''$ and for a ring $15''<R<60''$. The
central region is highly incomplete. Therefore the analysis of the
star counts in the seeing limited data was limited to ${\rm
mag_{K}}\leq16$ and $R>15''$.

\begin{figure*}[!p]
\centering
\includegraphics[angle=0,width=\textwidth]{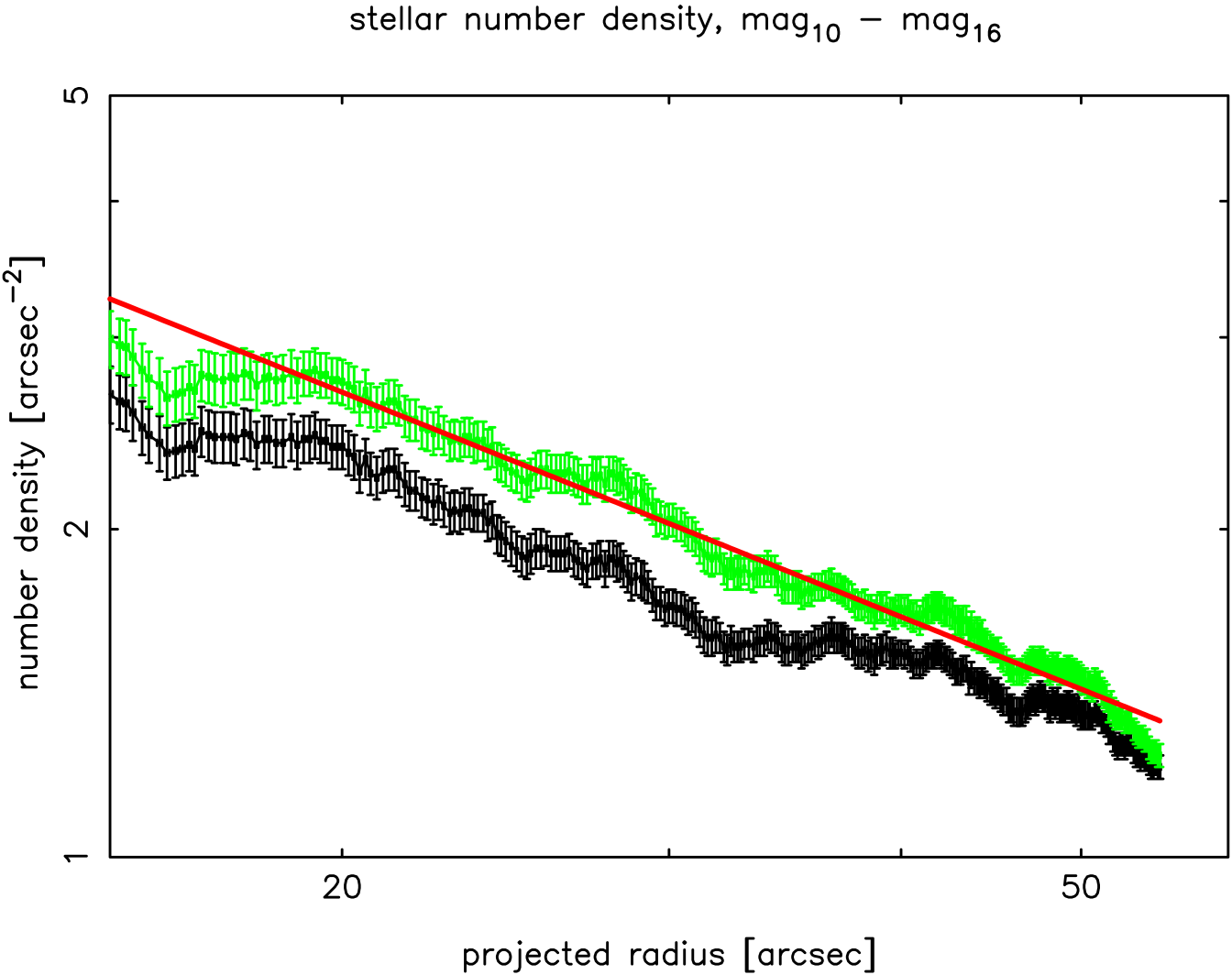}
\caption{Stellar number density extracted from ISAAC imaging
  observations at $2.09$\,$\mu$m for $\rm mag_{K} \leq 16$. The x-axis
  starts at $15''$. The lower, black data are completeness corrected
  counts, the upper green (grey) data have been additionally corrected
  for extinction. The straight line is a fit with a single
  power-law. It has an index of $\Gamma=0.68\pm0.02$.
\label{Fig:ISAAC_density}
}
\end {figure*}

In order to obtain the stellar number density, the stars were counted
in rings of variable width around Sgr~A*. Since the stellar density
decreases from Sgr~A*, it is desirable to vary the binning radius in
order to give each data point roughly the same statistical weight. The
radius of a circle that contained a given number of stars (in this
case 25) was determined for each pixel in the image. The azimuthal
average of this smoothing radius at each projected distance was then
used as the width of the rings around Sgr~A* in which the stellar
density was measured. The completeness and extinction corrected
stellar density for ${\rm mag_{K}}\leq16$ and $15''<R<60''$ obtained
in this way is shown in Fig.~\ref{Fig:ISAAC_density}.  The
uncertainties were calculated from the uncertainties of the
completeness correction, the masked area, and of the extinction
correction.  Error bars on the x-axis were estimated from the standard
deviation of the stars within a given ring from the mean radius of the
ring. For the extinction corrected data we added the uncertainty of
the radial location of the extinction value (because of the limited
resolution of the extinction map) quadratically to the uncertainty of
the error bars.

An excellent fit is provided by a power-law with an index
$\Gamma=0.68\pm0.02$, where the uncertainty is the formal fit
error. If the data are not corrected for extinction, the resulting
power-law index is $\Gamma=0.57\pm0.01$.

\subsubsection{The background light density \label{sec:bglight}}

Star counts on the ISAAC images have the limitation that the crowding
in the field is high and that the spatial resolution of the data is
limited by seeing. Due to the masking of regions of high
incompleteness, the analysis could not be extended to the central
$15''$. Also, it was limited to stars ${\rm mag_{K}}\leq16$. In order
to further constrain the structure of the faint, unresolved stellar
population, we analysed the background light density. Such a study was
pioneered by \citet{Philipp1999A&A} on the inner 30\,pc of the bulge,
but with a spatial resolution that was a factor $\sim$2 lower than in
the present work.

\emph{StarFinder} extracts point sources by PSF fitting, calculates a
smooth background, and provides images of the detected point sources,
of the background, and of the residuum. For background estimation, the
image is partitioned into rectangular sub-frames. The sky background
value computed for these sub-frames (via the IDL library routine SKY
that is based on a DAOPHOT routine of the same name) is assigned to
grid points corresponding to the centres of the frames. A smooth
background is then obtained by interpolation. An crucial parameter is
the size of the sub-frames, because the sub-image must contain, on the
one hand, a sufficiently large area for an accurate estimation of the
background, but, on the other hand, be small enough in order to
provide the required spatial resolution in case of a variable
background. We chose the default value provided by the
\emph{StarFinder} routine for our images, 22~pixels or
$2.94''$. Varying this parameter by $\pm1''$ had no significant
influence on the results.

\begin{figure*}[!p]
\centering
\includegraphics[width=\textwidth]{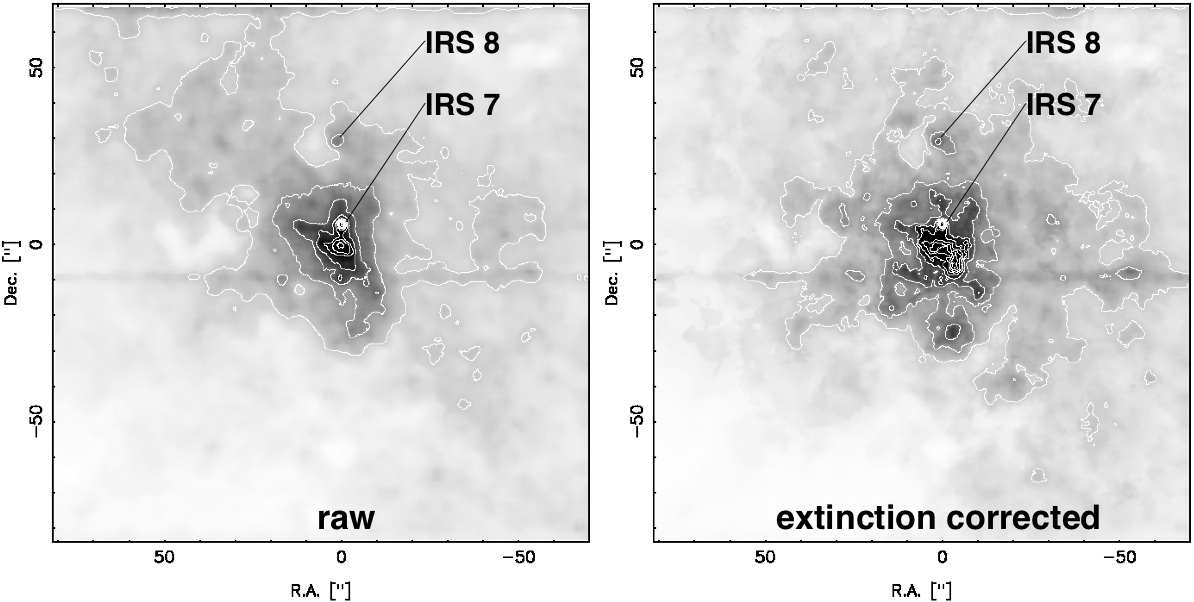}
\caption{Unresolved background light of the ISAAC $2.09$\,$\mu$m
  image. Left: Raw, uncorrected background as extracted with the
  \emph{StarFinder} code. Right: The background flux corrected for
  extinction (see Fig.~\ref{Fig:ISAAC_extinction}). The horizontal
  line in both images is an artefact due to a detector artefact in the
  original image that could not be removed. The brightest star in the
  field, IRS~7, and the bright, extended source IRS~8 produced
  artefacts in the background light. Contour lines are plotted in
  steps of 10\% from 10\% to 100\%, where 100\% corresponds to the
  maximum flux in the image. Darker shades correspond to higher flux
  densities.
\label{Fig:background}
}
\end {figure*}

Images of the background light density and of the extinction (see
\ref{sec:ISAAC_ext}) corrected background light density extracted in
this way are shown in Fig.~\ref{Fig:background}. The horizontal bar
that can be seen near the middle of both images is an artefact of the
image reduction process and is due to a detector feature that could
not be removed (it is also slightly visible in
Fig.~\ref{Fig:ISAAC}). The uncorrected background light density of the
cluster shows a clear extension along the direction
southwest-northeast, in agreement with the trend of the
extinction. The extinction corrected light density shows a more
symmetrical shape of the unresolved background. The brightest star in
the field, IRS~7, and the extended source IRS~8 could not be entirely
removed from the background and are still visible in the images of the
background flux. They were masked in the subsequent analysis. The
maximum of the background light before extinction correction (after
masking of the residual from IRS~7) is located $\sim$0.15$''$ west and
$\sim$0.6$''$ south of Sgr~A*. In the map of the extinction corrected
background light, the maximum is not located near Sgr~A*, but offset
$\sim$5$''$ to the southwest. This is due to the high correction
factor introduced by a region of apparently high extinction in this
region (an almost north-south running feature, visible in the left
panel of Fig.~\ref{Fig:ISAAC_extinction}). Since this feature is not
reproduced in the extinction map derived from NACO intermediate-band
imaging (see Fig.~\ref{Fig:extNACO}), we consider it an artefact and
mask this region in the analysis of the azimuthally averaged
background light density. A possible explanation for this artefact is
the presence of diffuse emission due to warm dust in the IRS3E-IRS~2
region \citep[see Fig.~\ref{Fig:outflow} and \ref{Fig:imbgresid} and
  also ][]{Moultaka2005A&A} that leads to very red colours of the stars
there.  After masking of this feature, the maximum light density after
applying the extinction correction remains centred on Sgr~A*.

\begin{figure*}[!p]
\centering
\includegraphics[width=\textwidth]{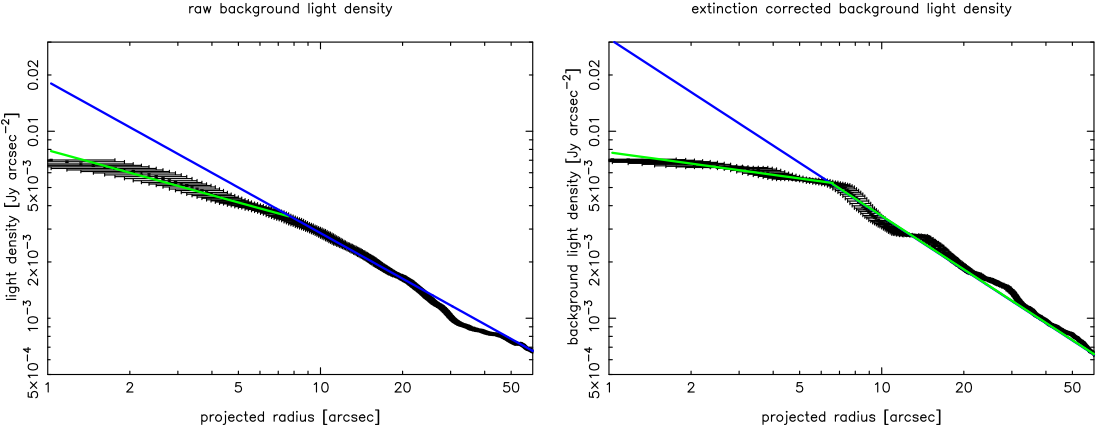}
\caption{Azimuthally averaged background light density in the ISAAC
  $2.09$\,$\mu$m image. Left: Raw light density. Right: Extinction
  corrected light density. The units are counts/pixel. The straight
  lines indicate single and broken power law fits. The parameters of
  the fits are $\Gamma_{\rm cusp}=0.40\pm0.05$, $\Gamma_{\rm
  cluster}=0.8\pm0.02$, and $R_{\rm break}=7.5''\pm1.0''$ for the
  uncorrected light density. After correction for extinction, the
  best-fit values are $\Gamma_{\rm cusp}=0.2\pm0.05$, $\Gamma_{\rm
  cluster}=0.95\pm0.03$, and $R_{\rm break}=6.5''\pm0.5''$.
\label{Fig:lightdens}
}
\end {figure*}

\begin{figure*}[!p]
\centering
\includegraphics[width=\textwidth]{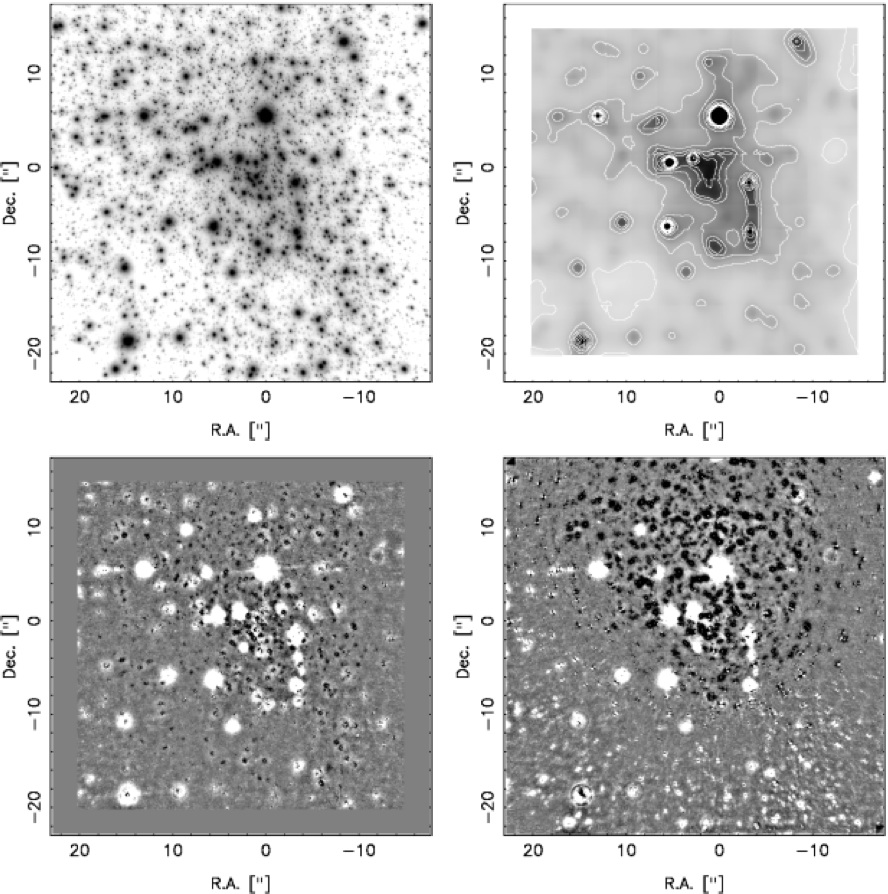}
\caption{Source extraction with \emph{StarFinder} from AO images. This
figure Illustrates the influence of bright stars on the determination
of the background and the effects of the spatially variable PSF. Upper
left: original image (2.27\,$\mu$m). Upper right: Background estimated
by \emph{StarFinder}, using a spatially variable PSF. Darker shades
correspond to higher flux densities. Lower left: Residual image when
using a spatially variable PSF.  Dark shades correspond to negative
residuals. Lower right: Residual image when using a single PSF.
Dark shades correspond to negative residuals.
\label{Fig:imbgresid}
}
\end {figure*}

\begin{figure*}
\centering
\includegraphics[angle=270,width=14cm]{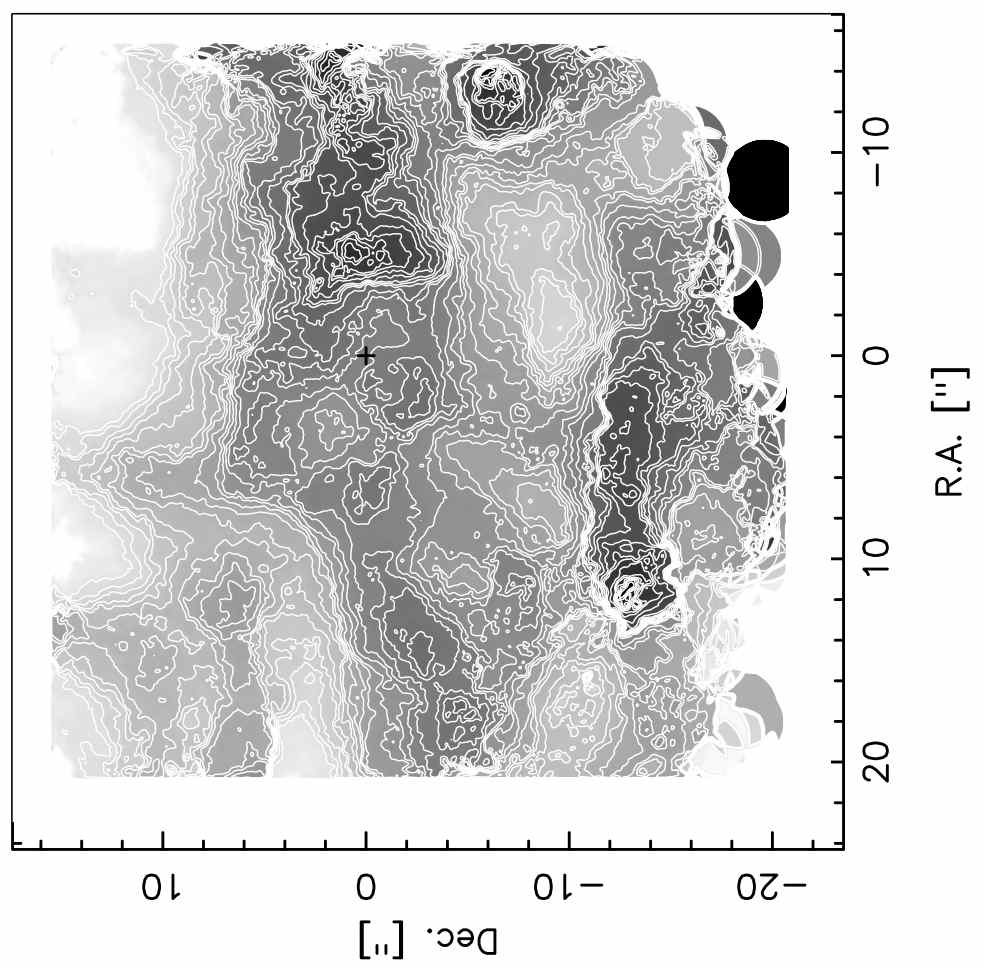}
\caption{Map of the interstellar extinction at $2.157$\,$\mu$m derived
  from NACO intermediate band imaging. The contours are spaced from
  $\rm A_{2.157\mu m}=2.0-3.5$\,mag at intervals of $0.05$\,mag.
  Darker shades correspond to higher extinction. The cross marks the
  position of Sgr~A*, where A$_{2.157\mu m}=2.6$\,mag.  The
  measurement at each pixel position results from the median of
  several tens of measurements within a projected distance of
  $2''$. The resolution of the extinction map is roughly ~$2''$.
\label{Fig:extNACO}
}
\end {figure*}

The azimuthally averaged background light density in rings of one
pixel width is shown in Fig.~\ref{Fig:lightdens}. The left panel shows
the uncorrected light density and the right panel the light density
corrected for extinction (scaled to the same values as the uncorrected
light density). The uncertainty of the average counts results from the
uncertainty of the mean of each ring. The main source of uncertainty
is the projected distance from Sgr~A* because the maps of the
background light density have only a limited spatial resolution. The
horizontal error bars in the figure correspond to the
1\,$\sigma$-width of a Gaussian function with FWHM equal to the
half-width of the sub-frames that were used to derive the background
light density. In case of the extinction corrected background light
density, the uncertainty due to the low spatial resolution of the
extinction map was added quadratically to the horizontal error bars. 

Single and broken power-laws were fit to the data, taking into account
the uncertainties in both axes. The best-fit results are shown in
Fig.~\ref{Fig:lightdens}. Overall, a broken power law provides better
fits to the data. The data at distances $\lesssim$7$''$ from Sgr~A*
prove that a single power law is not a good choice. The formal
1\,$\sigma$ uncertainties of the parameters for the power-law index of
the cusp, $\Gamma_{\rm cusp}$, of the overall cluster $\Gamma_{\rm
cluster}$, beyond the break radius , and the break radius $R_{\rm
break}$ are $\Gamma_{\rm cusp}=0.40\pm0.05$, $\Gamma_{\rm
cluster}=0.8\pm0.02$, and $R_{\rm break}=7.5''\pm1.0''$ for the
uncorrected light density. After correction for extinction, the
best-fit values are $\Gamma_{\rm cusp}=0.2\pm0.05$, $\Gamma_{\rm
cluster}=0.95\pm0.03$, and $R_{\rm break}=6.5''\pm0.5''$.

We tested two main sources of systematic uncertainties, that is, a)
possible bias introduced by an incorrect subtraction of the sky
background during data reduction and b) the accuracy with which
\emph{StarFinder} can extract the background flux from the crowded
stellar field. Tests were performed with different values for the
subtracted sky background.  These tests showed that the uncertainty of
the sky background is negligible for the slope of the cusp and the
value of the break radius. It contributes, however, a 1\,$\sigma$
uncertainty of $\leq 0.1$ to the slope of the stellar cluster beyond
the break radius. In order to test the second source of systematic
uncertainty, we produced images with a given, artificial broken power
law background structure and added  them to the images of the stellar
sources and of the residual provided by {\it StarFinder}. The
background was re-extracted from these artificial images and the
parameters of broken power-law fits to the azimuthally averaged light
density were determined. The 1\,$\sigma$ uncertainties estimated from
30 runs (a larger number is problematic due to the high demand of
computational time) are $\Delta a_{\rm cusp} = 0.1$, $\Delta a_{\rm
cluster}=0.1$ and $\Delta R_{\rm break} = 1.0''$.

The background light density provides a strong argument why the
power-law index of the stellar cluster \emph{must} decrease toward
Sgr~A*. From the single power law fits of the light density, we can
estimate that the surface brightness of the diffuse light near Sgr~A*
-- i.e.\ after subtraction of the stellar sources-- would be $\rm
mag_{K}\approx13.5$ per pixel at $R=0.1''$ and as high as
$mag_{K}\approx11.5$ per pixel at $R=0.01''$. This is clearly not
observed because these values are significantly higher than the
brightness of the actually observable stars within $\sim$$0.5''$ of
Sgr~A*, the so-called 'S-stars', which have K-band magnitudes
$\geq14$.

\subsection{AO observations}

\subsubsection{Source extraction}

 The great advantage of imaging observations with adaptive optics is,
of course, the high spatial resolution, needed to disentangle the
crowded point sources in the GC stellar cluster within a few seconds
of arc of Sgr~A*. However, there are also drawbacks of AO
observations, primarily caused by the PSF halos due to incomplete
correction and the limited size of the isoplanatic patch
($\sim$\,$15''$). This leads to a PSF that varies across the
field. The general consequence is a reduced photometric
accuracy. Deconvolution may give rise to numerous spurious sources
around brighter stars in the extended wings of their PSFs and in the
diffraction spikes due to the spider holding the secondary mirror of
the telescope.

The approach chosen in this work was to divide the AO mosaic image
into numerous overlapping sub-frames of size $10.8''\times10.8''$.
Source detection and photometry was conducted with \emph{StarFinder}
on these subframes. Since the size of the sub-frames is smaller than
the isoplanatic angle, this corresponds to PSF fitting with a
spatially variable PSF. The uncertainty of the photometry can be
estimated from the measured fluxes of stars that are present in
several of the overlapping fields. It was found that the relative
photometric uncertainty was $<10\%$ for more than $86\%$ of the stars.

Figure~\ref{Fig:imbgresid} shows the mosaic of the observations with
the intermediate band filter at $2.27$\,$\mu$m in the upper left
panel. In the upper right panel, it shows the background light density
determined with \emph{StarFinder}. Clearly, residuals related to
bright stars are present in the background. The lower left and lower
right panels show the residuals after PSF fitting and background
determination. The lower left panel shows the residual image when
applying a spatially variable PSF, the lower right panel shows the
residual image when a single PSF -- determined from bright stars
distributed across the entire FOV-- is used for the entire image. When
only a single PSF is used, there is a clear systematic effect visible
in the residuals: Near the guiding star, IRS~7, the residuals are
negative, while they become positive at larger distances.

For our analysis of the stellar number counts, we averaged the NACO AO
images from the observations at $2.27$\,$\mu$m and $2.30$\,$\mu$m.

\subsubsection{Extinction}

NACO intermediate band images at $\lambda=2.00, 2.06, 2.24$ and
$2.27$\,$\mu$m were used to derive the interstellar extinction.  Late-
and early-type stars do not have any absorption or emission features
at these wavelengths that would significantly affect the photometry in
the IB filters with their relatively broad $\Delta\lambda$. Minor
pollution by the CO band head in the $2.27$\,$\mu$m filter, by the
$Br\gamma$-line in the $2.24$\,$\mu$m filter, and by the HeI-line at
$2.058\,\mu$m is possible, but would only account for errors of a few
percent.

Since there were not sufficient calibration data available for an
accurate \emph{absolute} calibration of stellar fluxes in the
intermediate-band filter images, we applied the following method for
relative calibration of the fluxes in the different bands.  Fluxes at
each filter were calibrated using the stars S1, S2, S8, IRS~33N, and
W10 which are known as early type stars from spectroscopic
measurements by \citet{Eisenhauer2005ApJ} or
\citet{Ott2004PhDT}. Rayleigh-Jeans-like spectra could be assumed for
the calibration sources.  We assumed an extinction value toward the
calibration sources of $A_{K}=2.8$\,mag at $2.157$\,$\mu$m \citep[the
value given for S2 in][]{Eisenhauer2005ApJ} and an extinction law
$A_{\lambda}\propto\lambda^{-1.75}$ (see sec.~\ref{sec:ISAAC_ext}).

The detected stars can be approximated with a blackbody in the
observed wavelength range. Interstellar extinction influences the
observed spectral energy distribution (SED) of each star via
reddening. The colours of the stars between $2.00$\,$\mu$m and
$2.27$\,$\mu$m were determined by fitting a line to the measured
magnitudes at $2.00$, $2.06$, $2.24$, and $2.27$\,$\mu$m.  From the
assumed extinction law it was then straightforward to calculate the
interstellar extinction at $\lambda=2.157$\,$\mu$m toward the
individual sources. Stars with a blue SED were excluded from the
measurements because they were considered foreground stars.

In this procedure, the intrinsic colours of the stars have to be
known. In this work, only stars with apparent magnitudes $\rm mag_{K}
\leq 17.75$ were considered.  At the GC, main sequence stars of this
brightness range can only be O- or B-type stars. About a dozen B-type
main sequence stars have been identified near Sgr~A*
\citep{Eisenhauer2005ApJ}. Additionally, there are about 40
spectroscopically identified OB supergiants in the central
half-parsec, and a total of maybe 80 massive, young stars
\citep[outside of the central arcsecond][]{Paumard2006ApJ}. The vast
majority of the stars in the FOV covered by our observations, however,
are cooler giants or supergiants. We therefore assumed a blackbody
temperature of 5000\,K for all stars, except for the known early-type
stars \citep[taken from][]{Paumard2006ApJ}, for which we assumed
$T=20000$\,K. An error of 1000\,K in the temperature of a cool star or
of 10000\,K in the temperature of a hot star will lead to an error of
just $0.1$\,mag in the derived extinction at $2.157$\,$\mu$m.

In order to keep the uncertainty of the derived extinction low, the
median extinction was calculated in circles with $2''$ radius around
each pixel in the map. Thus a single extinction value was based on
50-120 measurements, except in the regions with the highest extinction
where just slightly more than 20 stars could be detected in such a circle.

The map of the interstellar extinction toward the GC determined in
this way is presented in Fig.~\ref{Fig:extNACO}.  Circular shapes near
the edges of the field are artefacts due to edge effects.  The
reference wavelength for the extinction map is $2.157$\,$\mu$m. The
statistical uncertainty at each point in the map is $<0.1$\,mag,
except in the regions of highest extinction, where the uncertainty is
$<0.2$\,mag. The extinction measured at the position of Sgr~A*, which
is marked by a cross in the figure, is A$_{\rm 2.157
\mu\,m}=2.6$. Compared with the value of 2.8\,mag at the position of
S2 \citep{Eisenhauer2005ApJ} this indicates a possible systematic
error of 0.2\,mag.

When comparing our results with the results of
\citet{Scoville2003ApJ}, we find generally good agreement. While with
our method, we cannot constrain exactly the absolute value of the
extinction without assuming an extinction law, its advantage is that
it is not dependent on the presence of line emission in the ISM and
provides thus a fairly homogeneous density of measurements across the
field. Also, we are not hindered by the brightness of Sgr~A* at radio
wavelengths. The agreement with the extinction map derived from the
seeing limited ISAAC images (Fig.\ref{Fig:ISAAC_extinction}) is also
favourable. The intermediate-band method with the NACO images is
certainly the more exact method. Therefore, in the analysis of the
ISAAC images, we chose to mask the elongated feature about $8''$
southwest of Sgr~A* that was apparent in the ISAAC extinction map (see
section~\ref{sec:bglight}). It is not seen on the NACO map.

\subsubsection{Completeness correction}

With NACO at the 8\,m-class VLT, imaging of isolated point sources as
faint as K$\sim$19 is easily achievable with integration times of the
order of several tens of seconds. However, when observing the central
parsec of the Milky Way, one is confronted with an extremely dense
stellar cluster and a large dynamic range due to numerous bright
sources distributed across the field: The brightest star, IRS~7, has a
K-magnitude of $\sim$$6.5$ and is heavily saturated on all AO
images. In addition, there are several dozen sources with a
K-magnitude around~10. Also, the field becomes considerably more
crowded near Sgr~A*. The detection efficiency for faint sources is
therefore highly variable (0-100\%) across an image. These effects
have to be taken into account by determining correction factors that
have to be applied to star counts in a given region for a given
magnitude.

\begin{figure*}
\centering
\includegraphics[width=14cm]{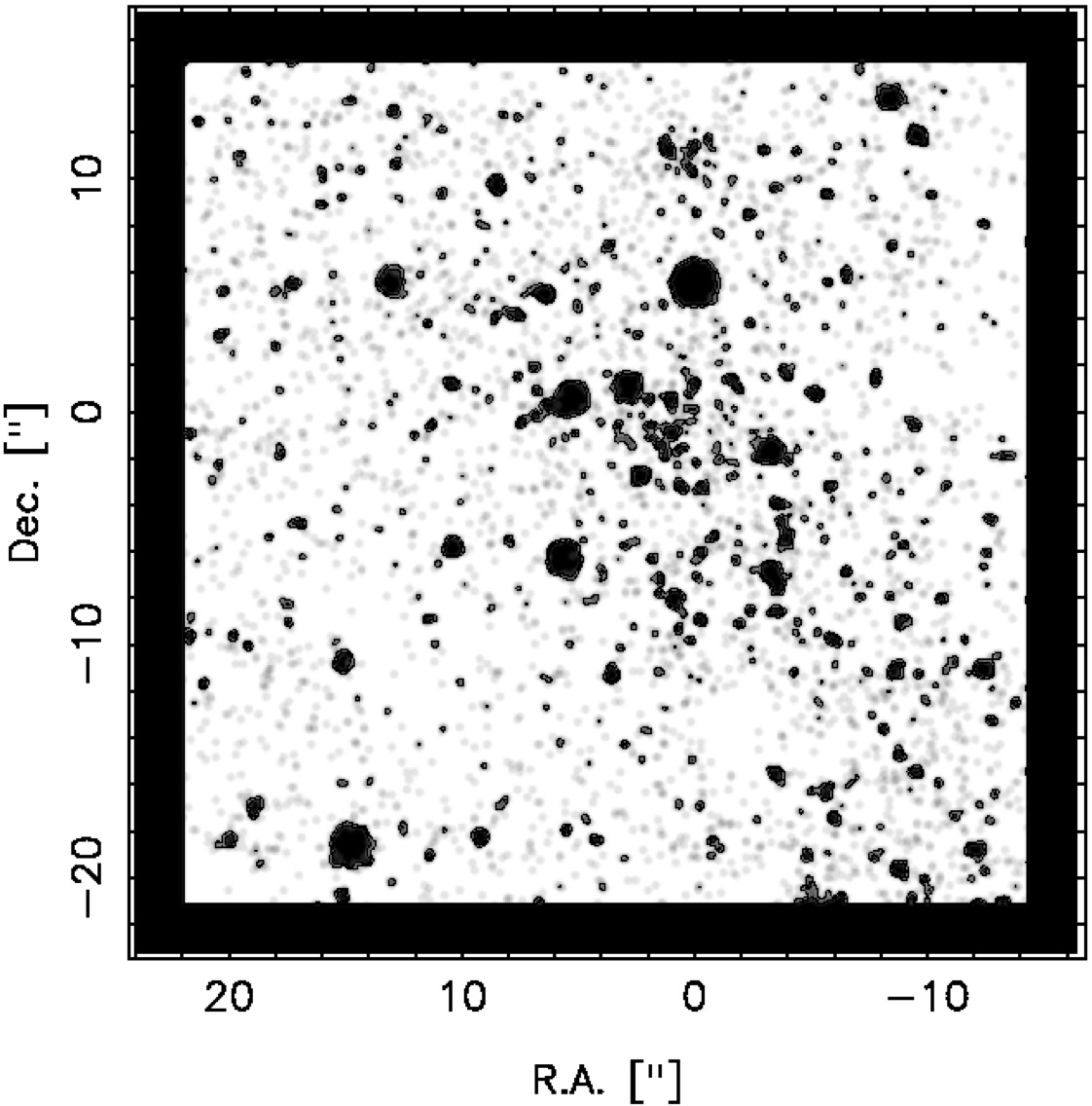}
\caption{Completeness map of the combined NACO $2.27+2.30\,\mu$m image
  for $\rm mag_{K}=16$. Completeness increases from darker to lighter
  areas. Contours are plotted at 30\% and 60\% completeness.
\label{Fig:completeness}
}
\end {figure*}

For determining the incompleteness of the images due to crowding and
due to the presence of bright stars we used the technique of adding
and recovering artificial stars \citep[see also][]{Genzel2003ApJ}. A
properly scaled PSF was added to the original image on grid points
spaced $0.5''$ apart. Subsequently, point source extraction was
performed with \emph{StarFinder} and the fluxes and positions of the
recovered stars were compared with the artificial input stars to see
whether a source could be recovered. By shifting the grid and
repeating the procedure various times, each image was effectively
probed with artificial stars on a $0.1''\times0.1''$-grid (The fine
sampling cannot be applied in one step because this would artificially
increase crowding). Maps containing the recovered artificial stars (as
delta functions, i.e.\ as pixels of value 1) were divided by maps
containing the input artificial stars (as delta functions) to
determine the percentage of completeness in the various regions of the
images (both maps of delta functions were convolved beforehand with
circles of $\sim$0.1$''$ radius in order to convert the discrete
sampling into continuous maps).  We created final maps of the
completeness for each magnitude by replacing the completeness at each
given pixel in the maps by the average completeness in a
$0.25\times0.25''$ box centred on that pixel. The uncertainty of the
completeness at a given pixel was then taken as the standard deviation
in this box. The completeness map for $\rm mag_{K} = 16$ is shown in
Fig.~\ref{Fig:completeness}. Table~\ref{Tab:NACO_completeness} lists
the completeness levels inside and outside of a projected radius
$R=8''$ from Sgr~A*.

The completeness correction was applied by dividing the delta maps
created from the star counts by the square root of the completeness
map (a number between 0 and 1) for the corresponding magnitude. The
square root was used in order to take into account the Poisson error
of the measurements and therefore the increased uncertainty of the
correction in areas of low completeness.  Areas where the completeness
was below 30\% were excluded from the analysis, i.e.\ masked in the
maps. Various values were tested for the completeness level where
masking was applied.  The values 20, 30, and 40\% gave similar
results, i.e.\ there is no strong sensitivity to the exact percentage
where masking is applied. Masking was also applied in a circular area
of $\sim$2$''$ radius centred on the heavily saturated star IRS~7.

\subsubsection{Luminosity function \label{sec:KLF}}

\begin{figure*}
\centering \includegraphics[width=\textwidth]{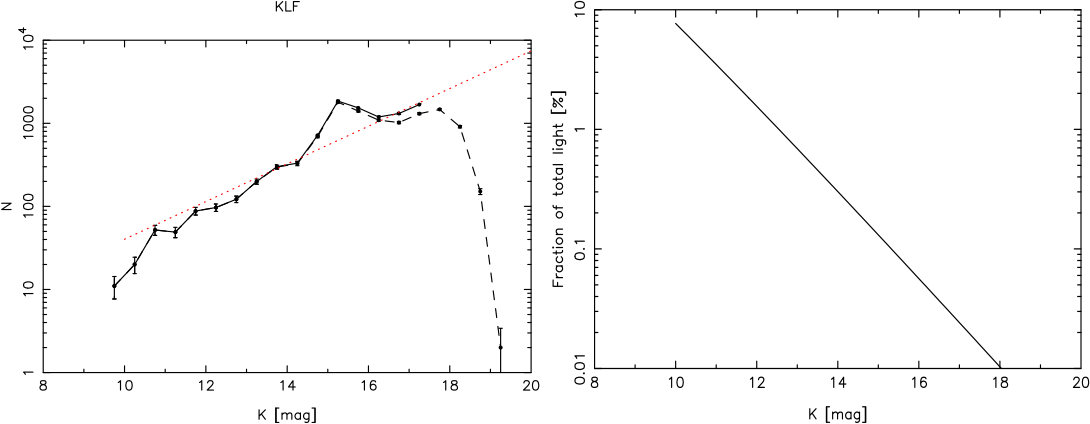}
\caption{The left panel shows the K-band luminosity function (KLF) of
  the GC stellar cluster as it was obtained from AO observations with
  NACO (field shown in Fig.~\ref{Fig:NACO}). The dashed line indicates
  the raw counts, the straight line the completeness corrected
  counts. The 'bump' at $\rm mag_{K} =15.25$ is due to red
  clump/horizontal branch stars.  The dotted line indicates a power
  law fit to the data -- ignoring the red clump bump --with a power
  law index $0.23\pm0.02$. The right panel illustrates the fraction of
  the total light of a star cluster that is contributed by stars of a
  given magnitude, under the assumption that the cluster LF is a pure
  power law. Here, the power law derived from the KLF in the panel
  above was used.
\label{Fig:LF}
}
\end {figure*}

The completeness corrected K-band luminosity function (KLF) is shown
in the left panel of Fig.~\ref{Fig:LF}. The dashed line is the raw LF
and the straight line the completeness corrected LF. The bump at $\rm
mag_{K} \approx 15.25$ is due to red clump (RC)/horizontal branch (HB)
stars. The dotted line indicates the power law index of the LF without
the HB bump. It is $\alpha=0.23$ with a formal fit uncertainty of
$0.02$. The right panel of Fig.~\ref{Fig:LF} illustrates in a
simplified way how much stars of a certain magnitude contribute to the
integrated light of the cluster, ignoring the HB bump and assuming
validity of the $\alpha=0.23$ power law index down to $\rm
mag_{K}=24$, which would correspond to main sequence stars of about
0.1~M$_{\odot}$ at the GC. This rough illustration shows that while
stars with $\rm mag_{K} > 16$ may make up more than 98\% of the stars
in the cluster, they contribute less than 1\% to the integrated light.

To first order, the effect of interstellar extinction is that the KLF
is shifted along the horizontal axis, depending on the value of
$A_{K}$.  The power law index of $\alpha=0.23$ determined from the KLF
was used in the subsequent analysis when correcting star counts for
extinction (see also sections~\ref{sec:ISAAC_counts} and
\ref{sec:bglight}).

The location of the KLF peak related to the RC is the reason why, in
this work, we chose to bin the stars into magnitude ranges
$13.75-14.75$, $14.75-15.75$, etc. Thus, the RC stars are largely
contained in a single bin.

\subsubsection{Two-dimensional stellar surface density \label{sec:densitymap}}

For each magnitude range, maps were created that contained pixel values
of 1, i.e. delta functions, at the position of a detected star.
Matching masks were created for each magnitude range for areas of
extremely low completeness. In these masks, the value of a pixel was
set to zero when the completeness at the corresponding position was
below 30\%. A list of all detected sources was created. If a star was
detected at a given distance, it was added to the list with its
position, magnitude, the completeness and extinction corrections at
the corresponding position, and a correction factor that takes into
account the masked area at the corresponding distance from Sgr~A* for
the corresponding magnitude of the star.

\begin{figure*}
\centering 
\includegraphics[width=\textwidth]{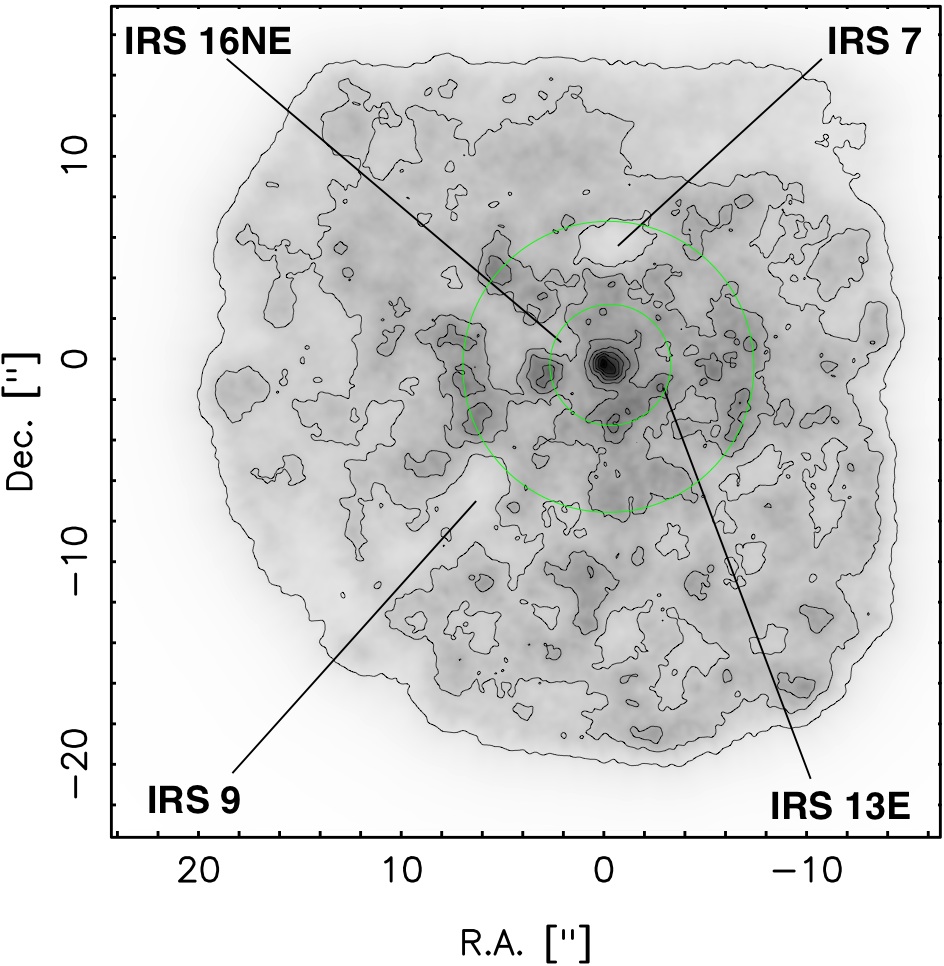}
\caption{Adaptively smoothed surface density map ($9.75 \leq \rm
  mag_{K}\leq17.75$) of the inner $\sim$$20''$ of the GC stellar
  cluster.  Contours are plotted in five steps of 0.67 times the
  maximum density. Darker shades correspond to higher surface
  densities. The adaptive smoothing was chosen such that 40~stars
  contributed to the density measurement at each pixel in the map. The
  resulting smoothing radius was $\sim$$0.6''$ at a distances $R<1''$
  from Sgr~A*, $\sim$$1''$ at $1''<R<10''$, and $\sim$$1.2''-2''$ at
  $R>10''$.  The map was corrected for extinction and
  completeness. Since regions of completeness $<30\%$ for given
  magnitudes were masked, the density near bright stars is
  under-estimated in this map. Some of these stars are indicated in
  the image. The masking is most probably the reason why the density
  peak is not centred on Sgr~A* because of the bright star IRS~16NW
  located $\sim$$1''$ north of Sgr~A*. The green (light gray) rings
  indicate projected distances of $3''$ and $7''$ from Sgr~A*. At
  these distances the structure of the cluster appears clumpy, with
  several density peaks superposed onto the generally smooth
  background. These structures are also present in the map before
  correction for extinction.
\label{Fig:dens2D}
}
\end {figure*}

In order to obtain a two-dimensional map of the surface density, an
adaptively smoothed map was created. The density at each position in
the map was obtained by dividing a constant number by the area of the
circle that contained this number of stars (after applying the
extinction and completeness corrections). The resulting map is shown
in Fig.~\ref{Fig:dens2D}. The number of stars that contribute to the
density measurement at each pixel was set to 40, providing a
reasonable compromise between high spatial resolution and suppressing
small-scale density fluctuations.  The necessary smoothing radius
varies between about $0.6''$ near Sgr~A* to about $1.5''$ at a
projected distance of $18''$ from Sgr~A*. Correction for masking is
difficult to achieve in a two-dimensional surface density map and was
therefore not implemented when creating the 2D-figures. Therefore,
some areas in Fig.~\ref{Fig:dens2D} show an artificially low surface
density. They are marked with the names of the associated bright stars
(e.g., IRS~7, IRS~16NE, IRS~9). Masking of IRS~16NW is probably
responsible for the asymmetry of the central cusp that appears
slightly offset to the south of Sgr~A*. Also, due to masking the
density shown in the map for the IRS~13E complex is certainly only a
lower limit.

The shape of the cluster appears close to circularly symmetric, but
not homogeneous. In Fig.~\ref{Fig:dens2D}, two circles are drawn
around Sgr~A* that mark projected distances of $3''$ and $7''$. The
$3''$ and $7''$ rings appear to contain some 'clumps', i.e.\
inhomogeneities of increased density.  One of these clumps is the
IRS~13E cluster, located at a projected distance of $\sim$3.5$''$ from
Sgr~A*.  This dense association of bright stars sharing a proper
motion in IRS~13E has received much attention
\citep{Maillard2004A&A,Schoedel2005ApJ,Paumard2006ApJ}.

\subsubsection{Radial profile of the number density}
\label{sec:density}

Theoretical expectations and our observational data (see
Fig.~\ref{Fig:background} and \ref{Fig:dens2D}) indicate that the
assumption of a spherically symmetric cluster is a good approximation
for the central parsecs. This is also supported by the distribution of
X-ray point sources \citep[see][]{Muno2006ApJS,Warwick2006JPhCS} and
large-scale near-infrared observations.  Plots of the azimuthally
averaged surface density were created by measuring the crowding and
extinction corrected counts in rings around Sgr~A*. The density was
determined in overlapping rings, with their mean distance increasing
in steps of one pixel. The width of the rings was chosen adaptively to
include 25 stars per smoothing circle, which translates to a variation
of the ring width from $\sim$$0.5''$ near Sgr~A* to $\sim$$1''$ at the
edge of the field. The value of the mean surface density was
determined for each ring. The corresponding uncertainty was calculated
from the uncertainties of the completeness correction, the masked
area, and of the extinction correction.  Because of the width of the
rings, the uncertainties of the average densities are fairly small.
The uncertainty in the projected distance that is related to each
density measurement was determined from the standard deviation of the
projected distances of all stars from Sgr~A* in the corresponding
ring.

\begin{figure*}
\centering
\includegraphics[angle=0,width=14cm]{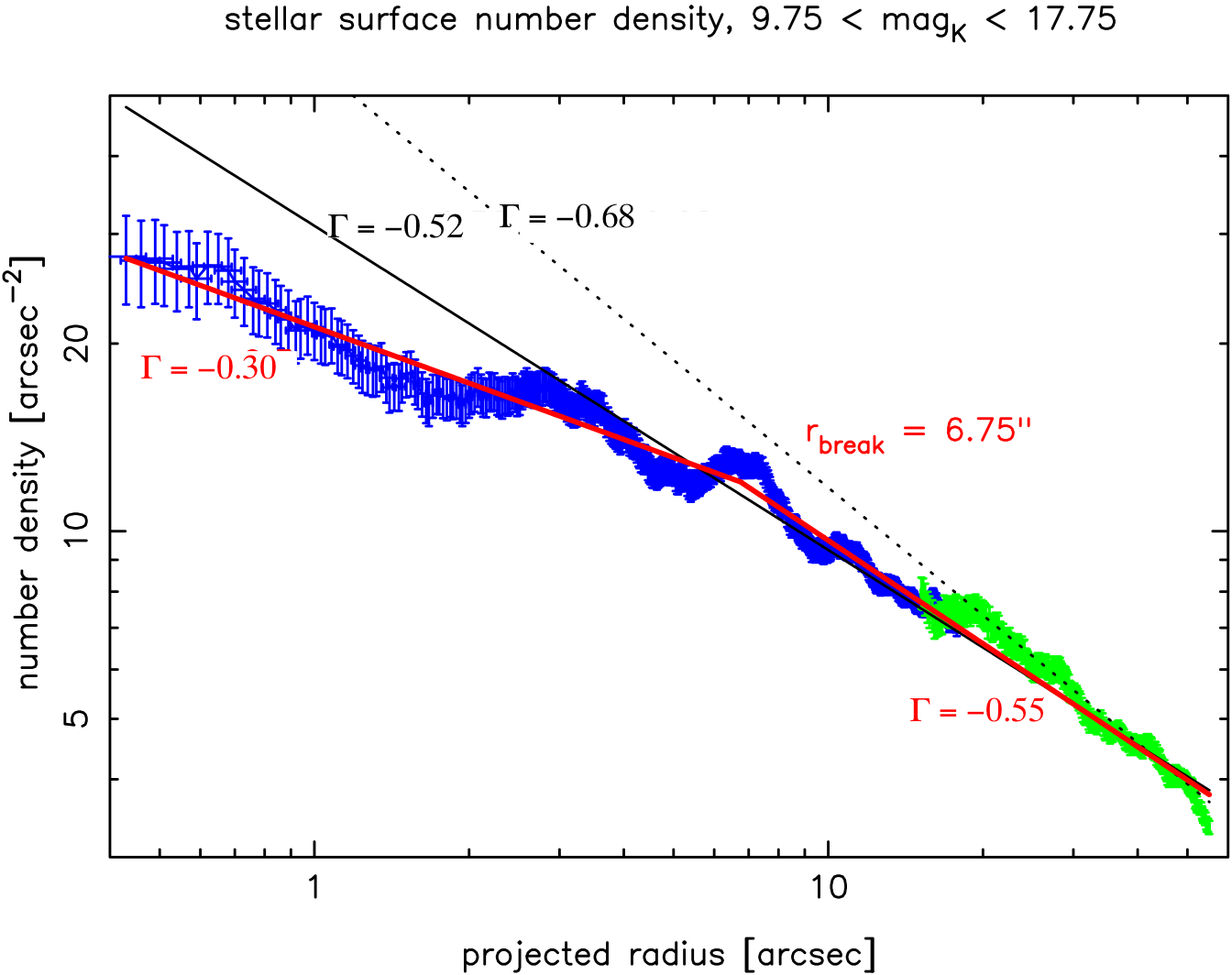}
\caption{Azimuthally averaged extinction and crowding corrected
surface density vs.\ distance from Sgr~A*. The green data points at
large distances are from the ISAAC star counts
(Fig.~\ref{Fig:ISAAC_density}). They have been scaled to the NACO data
(blue) by using the density in the overlapping region (The ISAAC data
are less deep than the NACO data). Fits to the NACO data with a single
(black) and with a broken (red) power-law are indicated along with the
corresponding power-law indices. The ISAAC data were not used for the
fits. The $1\sigma$ uncertainty of the power-law indices is $0.05$ and
of the break radius $0.05''$. The dotted line indicates the best-fit
power-law for the ISAAC data (see Fig.~\ref{Fig:ISAAC_density}).
\label{Fig:dens1017}
}
\end {figure*}

The plot of the surface density vs.\ distance from Sgr~A* for stars in
the magnitude range $9.75<\rm mag_{K}<17.75$ is shown in
Fig.~\ref{Fig:dens1017}. The straight lines illustrate fits with a
single and a broken power law. The dotted line indicates the power-law
fit to the ISAAC data alone (see Fig.~\ref{Fig:ISAAC_density}). It can
be seen that the broken power law provides a better fit than the
single power law. In a certain way this is to be expected because the
quality of a fit always improves with the number of
parameters. However, it can be seen that the single power law deviates
unacceptably from the data inside of about $2''$. This becomes
especially drastic if one considers the slope of the cluster far from
the cusp, as given, e.g., by the ISAAC data (dotted line in the
figure). Since it can be expected that the turnover into the shallow
cusp profile is gradual, it is not surprising that a shallower slope
of the large-scale cluster is measured when just using the NACO data.
The best fit values for the broken power law are a break radius of
$6.75\pm0.5''$, a power law index of $\Gamma_{\rm cusp}=0.30\pm0.05$
for the inner and $\Gamma_{\rm cluster} =0.55\pm0.05$ for the outer
part of the stellar cluster. The broken power law describes the shape
of the cluster remarkably well. Possible deviations from the assumed
broken power-law in form of apparent under- or over-densities may be
present at distances of about $3''$, $5''$, and $7''$.

\subsubsection{Star counts vs.\ light density \label{sec:countsvsdensity}}

Two recent publications on the structure of the central region of the
GC stellar cluster apply different methods that lead to different
results. On the one hand, \citet{Genzel2003ApJ} used surface number
density counts from AO observations with NACO at the ESO VLT and found
a power law cusp inside of $10''$ with an increasing stellar density
toward Sgr~A*.  \citet{Scoville2003ApJ}, on the other hand, used
Hubble space telescope data and analysed the diffuse light density of
the images by evaluating the median flux in circles around Sgr~A*,
only taking into account the pixels in the lower 80\% of the flux
distribution. They found a peak of the flux density at $1''-2''$ and a
drop inside of $1''$. As is shown in Fig.~\ref{Fig:LF} and noted in
section~\ref{sec:KLF}, the total light from the GC stellar cluster is
dominated by the contribution from the bright stars. The faintest stars
that can be discerned in Fig.~10 of \citet{Scoville2003ApJ} are
brighter than $\rm mag_{K}\approx15$. Assuming a simple power law LF
as in the left panel of Fig.~\ref{Fig:LF}, we find that the light from
stars fainter than $\rm mag_{K}=15$ contributes $\lesssim1.5\%$ of the
light, while stars brighter than $\rm mag_{K}=13$ contribute
$\sim$91\% of the total light. Due to the low spatial resolution of
the HST in the NIR this suggests that the diffuse light density in the
HST images is dominated by the brightest stars and the wings of
the PSF from the brightest stars, even if the brightest $20-50\%$ of
the pixels are discarded.

This hypothesis was tested in a simple way by creating artificial
images using the lists of identified stars in the high-resolution
NACO/VLT AO images combined with the appropriate point spread function
and the known magnitudes of the stars. Two maps were created, one
containing stars in the magnitude range $9.75-17.75$, the other one in
the range $14.75-17.75$. The surface density of the light was
determined in the same rings that were used to extract the number
density (see section.~\ref{sec:density} for details). Completeness and
extinction corrections were taken into account. Following
\citet{Scoville2003ApJ}, the light density was determined by taking
the median of the pixel brightness distribution in the rings and by
taking into account only the fainter 80\% of the pixels.

\begin{figure*}
\centering
\includegraphics[width=\textwidth]{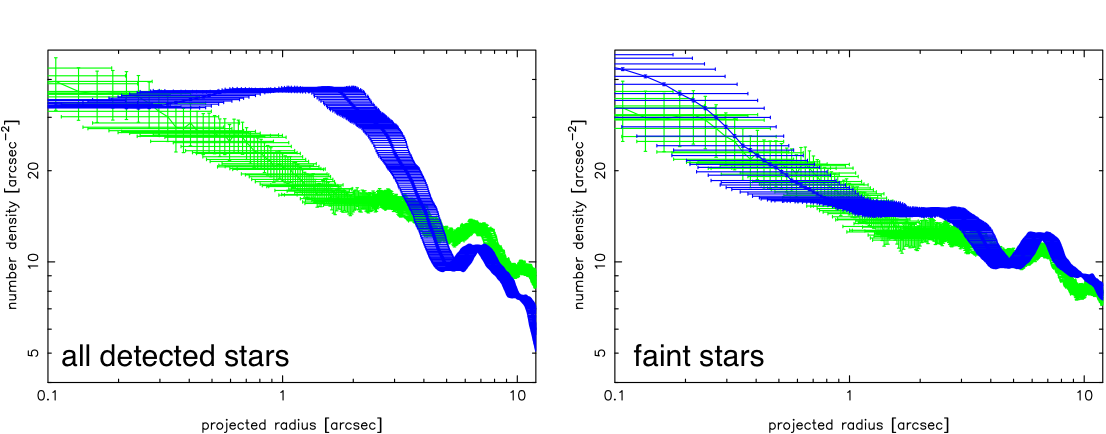}
\caption{Comparison between the extinction and crowding corrected
  surface number density (green) of stars in the magnitude ranges
  $9.75-17.75$ (left) and $14.75-17.75$ (right) and the corresponding
  light density (blue). The light density was derived from artificial
  images, taking into account only the fainter 80\% of the pixel
  brightness distribution.  The average light density was scaled to
  the average surface number density.
\label{Fig:light}
}
\end {figure*}

The results of this test are shown in Fig,~\ref{Fig:light}. In the
left panel, the radial light density extracted from the map with stars
$\rm 9.75\leq mag_{K} \leq 17.75$ is compared to the corresponding
number density (the average light density was scaled to the average
number density). In the right panel a similar plot is shown for stars
$\rm 14.75\leq mag_{K} \leq 17.75$. The plot that includes the bright
stars looks fairly similar to Fig.~12 shown in
\citet{Scoville2003ApJ}. The density maximum is reached at projected
distances of $1''-3''$ from Sgr~A*. The peak of the light density and
its decrease inside of $1''$ are less pronounced in
Fig.~\ref{Fig:light} than in the corresponding plot by
\citet{Scoville2003ApJ}. The probable cause is that in this work, we
only considered stars fainter than $\rm mag_{K}=9.75$. The star
IRS~16NE, with $\rm mag_{K}\approx9.0$ is located at a projected
distance of just $3.0''$ from Sgr~A*. It was not considered in our
test, but is present in the HST image. It is evident that the bright
stars have a significant influence on the surface light density. The
surface light density can therefore be misleading when determining the
shape of the GC stellar cluster. The excess light density at projected
distances $1''-5''$ from Sgr~A* is probably related to the presence of
bright, young stars in this region \citep[e.g.,][]{Paumard2006ApJ}.
In marked contrast, the trend of the azimuthally averaged light
density of the faint stars is very similar to the surface number
density, as expected for a homogeneous stellar population.

\begin{figure*}
\centering
\includegraphics[width=\textwidth]{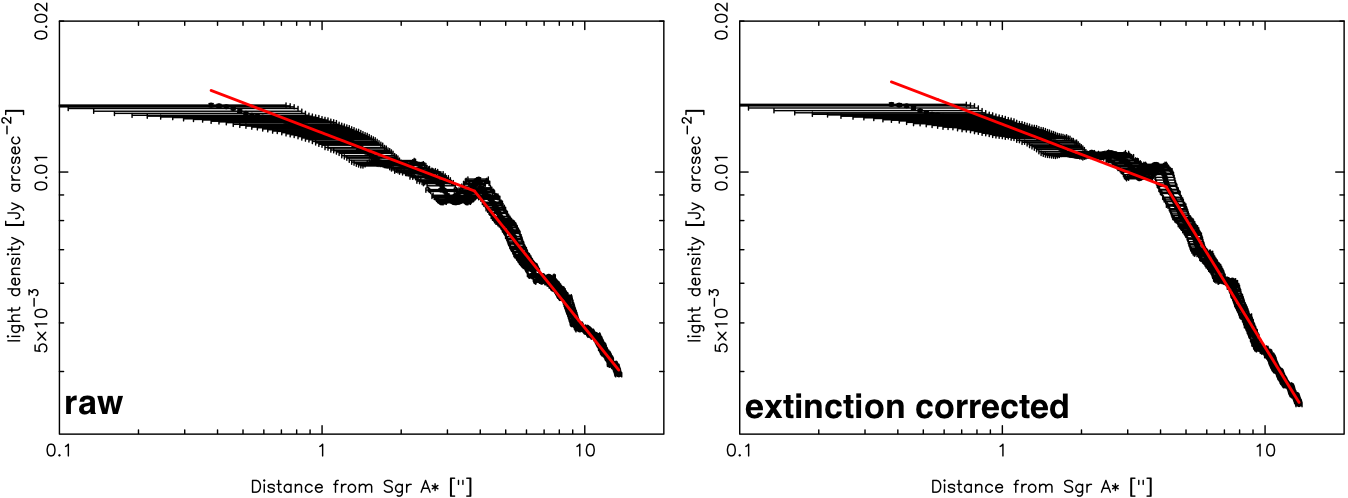}
\caption{Azimuthally averaged background light density
  extracted from the NACO AO data.  The raw light density is shown in
  the left panel, while the background corrected light density is
  shown in the right panel (normalized to the raw counts).  The
  straight red lines show a broken power law fit to the data. The
  parameters of the fit are $R_{\rm break}=4.0\pm0.8''$, $\Gamma_{\rm
  cusp} = 0.15\pm0.07$, and $\Gamma_{\rm cluster}=0.85\pm0.07$ for the
  extinction corrected data (uncorrected: $R_{\rm break}=3.8\pm0.8''$,
  $\Gamma_{\rm cusp} = 0.20\pm0.08$, $\Gamma_{\rm
  cluster}=0.65\pm0.10$).
\label{Fig:bgnaco}
}
\end {figure*}

The analysis of the seeing limited ISAAC data (see
section~\ref{sec:bglight}) shows that the diffuse light density can be
examined well by extracting the point sources via PSF fitting combined
with simultaneous background estimation. A particular difficulty with
AO imaging data is, however, the spatially variable PSF. Combined with
the extreme source density in the central arcseconds, this has the
effect that estimating the wings of the PSF is subject to possibly
significant uncertainties. This can lead to considerable residuals
present in the smooth, fitted background light density in the vicinity
of  brighter stars (see Fig.~\ref{Fig:imbgresid}, upper right
panel). By rejecting the brightest 50-80\% of the pixels, similar to
the analysis in \citet{Scoville2003ApJ}, however, the influence of
these residuals of the bright stars can be minimised when extracting
the azimuthally averaged light density. We applied this method and
show the resulting plots of the background light density vs.\ distance
from Sgr~A* in Fig,~\ref{Fig:bgnaco}. Both the raw and extinction
corrected background light density can be fitted well with a broken
power law. The parameters for the extinction corrected data are
$R_{\rm break}=4.0\pm0.8''$, $\Gamma_{\rm cusp} = 0.15\pm0.07$, and
$\Gamma_{\rm cluster}=0.85\pm0.07$. Here, the uncertainties include
the formal uncertainty as well as the uncertainty due to the
reliability of the extraction of the background light from the AO
images.  The latter was tested by creating artificial images with a
given background, and varying the break radius and the power-law
indices (as described for the seeing-limited data in
section~\ref{sec:bglight}). The background could be reliably
recovered.

\subsection{Adaptive-kernel estimates of the surface and space densities}

Estimating the surface density by counting stars in bins has a number
of disadvantages \citep[e.g.][]{Scott1992book}: the estimates are
discontinuous, the bias-variance tradeoff is far from ideal,
confidence intervals on the density are difficult to compute,
etc. Also, it is impossible in principle to differentiate a binned
density profile, which means that quantities like the space density
are impossible to compute unless parametric forms are fit.

All of these shortcomings can be addressed by adopting techniques from
nonparametric function estimation theory.  Here, we apply an adaptive
kernel method to estimate $\Sigma$ and $\rho$
\citep{Merritt1994AJ}. We imagine that each star on the 2D image is
replaced by a circularly-symmetric Gaussian kernel.  Averaging this
kernel over angle at fixed distance from the black hole generates a
1D, asymmetric kernel which can be used to compute an estimate of
$\Sigma(R)$:
\begin{eqnarray}
\Sigma(R) &=& \sum_{i=1}^N {w_i\over h^2} K(R,R_i,h),\\
K(R,R_i,h) &\equiv&{1\over 2\pi} e^{-(R_i^2+R^2)/2h^2}I_0(RR_i/h^2).
\end{eqnarray}
Here, $R_i$ is the projected radius of the $i$th star, $w_i$ is the
weight associated with the $i$th star, and $h$ is the kernel width,
roughly analogous to the bin width; $I_0$ is the modified Bessel
function.  We varied the kernel width as $R^{0.8}$, or roughly as
$\Sigma^{0.5}$; the mean value of $h$ was chosen so as to give roughly
the same degree of smoothing as in the binned estimates of Figs. 5 and
6.  Estimates of the space density $\rho(r)$ were obtained by applying
Abel's formula directly to the $\Sigma$ estimates, using a larger
value of $h$ in order to compensate for the increased variance due to
the differentiation.  Bootstrap confidence intervals on all quantities
were generated via resampling \citep{Efron1982book}.

Figure~\ref{Fig:kernel} shows the results, compared with the
$\Sigma\sim R^{-3/4}$ and $\rho\sim r^{-7/4}$ Bahcall-Wolf
predictions.  The mean indices are lower than predicted, and exhibit
what appear to be statistically significant wiggles. The general
structure of the cluster, i.e. a broken power-law with a break radius
around $7''$ and a rather flat cusp can be seen in the kernel
estimates as well.

\section{Discussion}

\subsection{Contamination by the discs of young stars \label{sec:contamination}}

Considering the one/two discs of young stars discovered and described
by \citet{Levin2003ApJ} and \citet{Genzel2003ApJ} and described in
detail by \citet{Paumard2006ApJ}, the question arises about the extent
by which the stellar surface density derived in this work is
contaminated by stars in these discs. It is not straightforward to
assess how many young stars contribute to the surface density at each
radius. However, we can give a rough estimate by using the information
in \citet{Paumard2006ApJ}. In their Fig.~13, they plot the luminosity
functions of the spectroscopically identified young stars. They find a
very flat slope, from which they conclude that the IMF of the stellar
discs must be much flatter than the slope of the Miller-Scalo IMF.
Extrapolating from their KLF between K-magnitudes 11 and 14 shows that
the surface density of faint ($K=16-17$) stars that may belong to the
young stellar discs should be only a factor 5-10 higher than the
surface density of the brighter stars. This would result in a surface
density of faint young stars of $\leq 2$\,arcsec$^{-2}$ within $2-5''$
of Sgr~A*.

Alternatively, \citet{Paumard2006ApJ} also show that the surface
density profile of the discs is very steep, with an exponent of about
$-2$ (see their Fig.~6). They have spectroscopically identified the
brighter stars with $\rm mag_{K} < 15$. Taking the density of the
discs between projected radii of $2''-5''$ and scaling it by factors
$\sim$10-20 in order to adjust it for the fainter stars \citep[using
the KLF of][]{Paumard2006ApJ}, surface densities of 1-2 young stars
arcsec$^{-2}$ within $2-5''$ of Sgr~A* are obtained. Therefore, since
the faint stars at magnitudes 15-17 dominate the star counts, the
surface density plots in Fig.~\ref{Fig:dens1017} should not be
contaminated significantly by the stars in the young stellar discs. An
exception are the stars with K-magnitudes between 14 and 15. As
\citet{Eisenhauer2005ApJ} have shown, stars in this brightness range
within about $0.5''$ of Sgr~A* appear to be B-type main sequence
stars.

The concentration of bright, young, massive stars in the stellar discs
within a few arc seconds of Sgr~A* is, however, reflected nicely in the
plot of the surface light density shown in
Fig,~\ref{Fig:light}. The light density shows a clear excess at
projected distances between about $1''$ and $4''$ from Sgr~A*.

\subsection{Mean stellar mass probed, mass segregation \label{sec:masses}}

Figure~\ref{f:meanM} shows the predicted K-band luminosity function
(KLF), the mean stellar age and the mean stellar mass for all stars
and the old ($>1$ Gyr) stars, as function of their K-band
luminosity. This population synthesis model for the central parsec
\citep{AlexanderSternberg1999ApJ,Alexander2005PhR} assumes a
continuous star-formation history over the lifetime of the Galaxy with
a "normal" initial mass function
\citep[here][]{MillerScalo1979ApJS}. The model is based on the
$Z=1.5Z_\odot$ stellar evolution tracks of \citet{Schaller1992A&AS}
and \citet{Girardi2000A&AS}. The continuous star formation history is
indicated by the very good match between this model and the observed
KLF (evolved red giants and massive blue giants) in the inner GC
\citep{AlexanderSternberg1999ApJ}, as well as by models of the
luminosity in the GC on the $\sim 50$ pc scale \citep{Figer2004ApJ}.
The model does not take into account the recent star formation
processes reflected in the two discs of young, massive stars in the
inner $10''$ \citep[e.g.\ ][]{Genzel2003ApJ}. However, this should
only affect significantly the bins brighter than $\rm mag_{K} \leq 13$
(see also discussion in section~\ref{sec:contamination}).

The mean mass is quite constant in the range $\sim\!13$--$17$ mag,
$\left\langle M_{\star}\right\rangle \!\sim\!3\, M_{\odot}$ (main
sequence life span $\sim$0.5\,Gyr), with a distinct drop to
$\left\langle M_{\star}\right\rangle \!\gtrsim\!2\, M_{\odot}$ at
$K\!\sim\!15.5$ mag, due to the concentration of red clump horizontal
branch giants (main sequence life span $>1$\,Gyr) in that magnitude
range, as observed in the measured KLF \citep[see Fig.~\ref{Fig:LF} in
this work and also Fig.~9 in ][]{Genzel2003ApJ}. The mean mass of the
entire population (including fainter stars and remnants) is
$\left\langle M_{\star}\right\rangle \!=\!0.5\, M_{\odot}$ \citep[][,
table 2.1]{Alexander2005PhR}. Thus, the mean star included in our
counts is expected to have an intermediate mass between the typical
numerous low mass objects, and the rarer massive stars and stellar
mass black holes, which probably dominate the very centre due to mass
segregation
\citep{Morris1993ApJ,Escoude2000ApJ,HopmanAlexander2006ApJ,Freitag2006ApJ}. In
the restricted $K$-range near the horizontal branch feature,
$15\!\lesssim\!  K\!\lesssim\!16$ mag, up to 80\% of the stars are
old, and are thus more likely to have attained their steady-state
distribution. The mean mass of the old stars is $\left\langle
M_{\star}\right\rangle \!\simeq\!1.5\, M_{\odot}$.  Mass segregation
substantially shortens the relaxation time below the typical value
estimated in section~\ref{sec:structure} because it increases the mean
stellar mass \citep[see e.g.\ section 3.2.1 in ][]{Alexander2005PhR}.
A numerical model of mass segregation in the GC
\citep{HopmanAlexander2006ApJ} indicates that the relaxation time
falls from $\sim 1$ Gyr at $0.4$ pc to $\sim 0.4$ Gyr (the typical
lifespan of a $3 M_\odot$ star) at $0.02$ pc. The HB/RC progenitors
($<2 M_\odot$) have lifespans $>1$ Gyr and are thus expected to be
well-relaxed and segregated. The dynamical state of the $\sim 3
M_\odot$ stars brighter and fainter than the RC giants is less
certain, since their lifespans are of the same order of magnitude as
the relaxation time, but they could plausibly be segregated to a
substantial degree. We thus expect stars in the 15-16 magnitude range
to have the shallowest slope in comparison with stars immediately
brighter and fainter.

Based on these theoretical predictions, we examined the azimuthally
averaged surface density for magnitude bins of $13.75-14.75$,
$14.75-15.75$ (HB/RC stars), and $16.75-17.75$. In the KLF shown in
the left panel of Fig.~\ref{Fig:LF} one can see that the intermediate
bin includes most HB/RC stars, while the bin with the faintest stars
should be largely free of HB/RC stars. The bin for the brighter stars
may contain some HB/RC stars, but was chosen because it contains a
statistically more significant number of stars than the next brighter
bin. The resulting number density plots for the different stellar
types, along with broken power law fits, are shown in
Fig.~\ref{Fig:densmag}. We find systematically different slopes on the
$\geq 3\sigma$ level.  Stars brighter or fainter than the HB/RC stars
have somewhat shallower power-law indices outside of
$\sim$10$''$. Possibly, this is due to star formation activity.  As
qualitatively anticipated by the stellar mass model
(Fig.~\ref{f:meanM}) and as predicted for a relaxed multi-mass cusp
\citep{BahcallWolf1977ApJ}, we find a shallower cusp for the low-mass
stars that dominate the bin mag$_K=15-16$. However, the slope of the
cusp shallower than predicted by the Bahcall-Wolf solution, and the
difference between this slope and that of the adjacent fainter and
brighter magnitude bins, which are dominated by more massive is much
larger than found in the numeric simulations of \citet{Freitag2006ApJ}
for $2 M_\odot$ and $3 M_\odot$. These deviations from the theoretical
predictions cast some doubt whether the observed trend in the cusp
slopes is due to mass segregation, or perhaps reflects different
dynamical histories for the old, low-mass stars and young and massive
ones.

There appears a curious feature in the surface density of the
intermediate-bright stars, a ``trough'' or under-density, at a
projected radius of about $5''$. It may be related to a lack of HB/RC
stars at this distance. The data and power-law fit presented in the
middle panel of Fig.~\ref{Fig:densmag} indicate that the feature is
clearly significant on the $>5\sigma$ level (very small error bars
would result when averaging the density between, e.g., $4''-6''$).
The dip around $5''$ is also present in the surface density plots
presented by \citet[][, their Fig.~4]{Eckart1993ApJ} (although with a
low significance) and \citet[][, their Fig.~7]{Genzel2003ApJ}. We do
not detect this feature in the surface density of the unresolved
background, which is expected to trace the stellar population with
K-magnitudes $>17-18$, and thus stellar masses comparable to or lower
than the HB/RC stars. We conclude that the feature is therefore due to
an under-density of HB/RC stars at this distance or -- alternatively
-- an over-density of similarly bright stars of a different, possibly
early type at slightly larger and smaller distances. Early type stars
could be related to the few million year-old starburst in the central
parsec \citep[e.g.,][]{Allen1990MNRAS,Krabbe1995ApJ,Paumard2006ApJ}.
\citet{Genzel2003ApJ} find a clockwise and a counter-clockwise
rotating ring of young stars in the central half-parsec with
characteristic radii of $2''-4''$ for the clockwise and $4''-7''$ for
the counter-clockwise one, respectively. If the systems of young stars
are ring-like with most stars concentrated at these radii then there
could exist an over-density of intermediate-bright stars at the
corresponding distances. In fact, as discussed in the section above,
the star counts in the brightness range of the HB/RC stars may be
contaminated in a non-negligible way by young stars in the innermost
arcseconds.  More detailed spectroscopic information is needed in
order to examine this feature of the cluster further.

\subsection{Clumpiness of the cluster}

The analysis of the surface density map of the GC stellar cluster
showed that besides the previously known density concentration of
stars in the IRS~13E cluster there appear to exist several other
significant localised concentrations or clumps of stars.  Recent
observations and theoretical work indicate that the formation of the
young stars in the GC has probably occurred in massive accretion discs
\citep[see][]{Nayakshin2006MNRASa,Nayakshin2006MNRASb,Paumard2006ApJ}. The
identification of dense groups of co-moving stars in the GC cluster
can then be interpreted in the sense that fragmentation of the discs
and subsequent star formation in these fragments may produce fairly
large sub-groups of stars. Highly clustered star formation in the
discs of young stars is indeed indicated in simulations (S.\
Nayakshin, private communication). Detailed spectroscopic information
and proper motion studies are needed in order to clarify the nature of
the clumps in the cluster.

\subsection{Extinction and structures in the ISM}

Figure~\ref{Fig:outflow} shows contours of the interstellar extinction
(see Fig.~\ref{Fig:extNACO}) superposed on a NACO L'-band
($3.8$\,$\mu$m) image.  The so-called mini-cavity to the southwest of
Sgr~A* can be distinguished clearly in this image and is indicated by
a white dashed line. Linear filaments, possibly shock fronts, mark its
western edge. It has been suggested that the mini-cavity might be
created by fast winds from the HeI-stars in the the central star
cluster or - more speculative- by an outflow from Sgr~A*
\citep{Lutz1993ApJ,Melia1996ApJ,Yusef-Zadeh1998ApJ}.

It is interesting to note that there is a narrow, channel-like feature
aligned southwest-northeast running right across Sgr~A* and toward the
mini-spiral. Cometary-like sources (indicated by blue arrows in the
Figure) are aligned with this structure and point right at
Sgr~A*. Mu\v{z}i\'{c} et al.\ (in prep.) discuss these features along
with filamentary structures and their proper motions in L'-band images
and relate them to a putative outflow from the central arcseconds.

In theoretical models that are used to understand accretion and
emission processes near Sgr~A*, so-called RIAF models
\citep[radiatively inefficient accretion flow, see,
e.g.,][]{Quataert2003ANS} the extremely low accretion efficiency is
almost inevitably related to the presence of outflows \citep[see also
the ADIOS model by][]{BlandBegel1999MNRAS}. Also, there a theoretical
and observational arguments that favour the presence of a jet in
Sgr~A*
\citep[e.g.,][]{Melia2001ARA&A,Markoff2001A&A,Yuan2002A&A,Bower2004Sci}.
As for the orientation of the outflow, \citet{Eckart2006A&Ab} and
\citet{Meyer2006A&Aa} note that the electric field vector during a NIR
flare was oriented along the direction southwest-northeast, similar to
the direction of the channel-like feature that is centred on Sgr~A*. A
very similar angle was found during a second NIR flare event
\citep{Meyer2006A&Ab,Trippe2006astroph}. In combination with the
theoretical considerations and the other evidence for an outflow from
the region near Sgr~A*, the low-extinction channel found by us may
represent a new piece of evidence for the existence of an (aligned)
outflow.

\subsection{Overall structure of the cluster \label{sec:structure}}

\begin{figure}
\centering
\includegraphics[angle =270,width=0.8\textwidth]{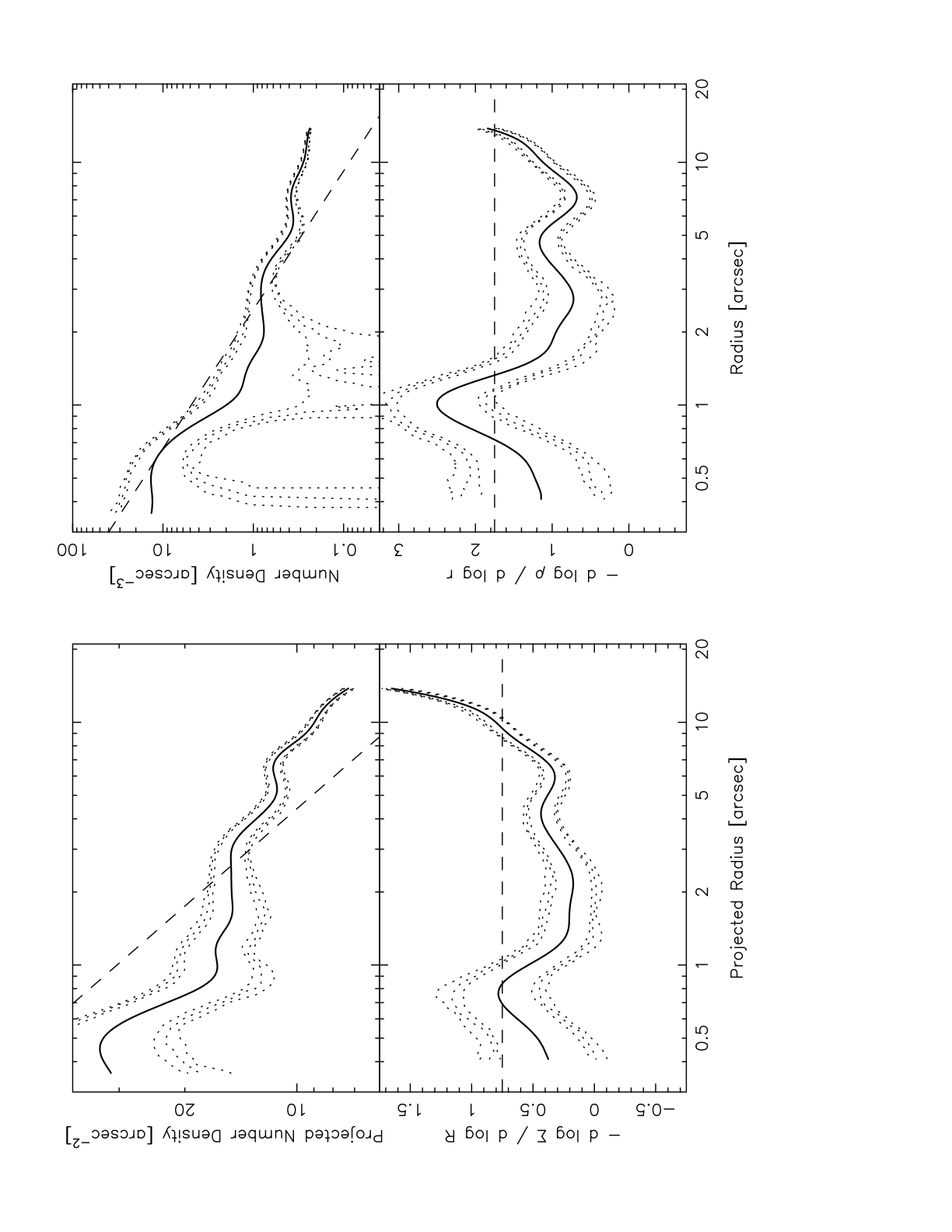}
\caption{The structure of the GC stellar cluster as derived from the
  NACO star counts ($9.75 \leq {\rm mag_{K}} \leq 17.75$) with the
  nonparametric kernel estimator \citep[see][]{Merritt1994AJ},
  including all relevant corrections. Upper
  panel: Non-parametric kernel fit of the star counts. The dotted
  lines indicate the 90, 95, and 98\% confidence limits from the
  bootstrap method described in \citep[see][]{Merritt1994AJ}. The
  dashed line indicates the slope of a Bahcall-Wolf cusp
  \citep{Bahcall1976ApJ,BahcallWolf1977ApJ}, i.e. $\sigma\propto
  R^{-0.75}$ for the surface density and $\rho\propto r^{-1.75}$ for
  the space density, where $R$ is the projected and $r$ the 3D
  distance from Sgr~A*. The lower panel shows a plot of the power-law
  slope of the stellar surface density vs.\ distance from Sgr~A* as it
  results from the kernel fit. The dashed line indicates the slope of
  a single-mass Bahcall-Wolf cusp.
\label{Fig:kernel}
}
\end {figure}

\begin{figure}
\includegraphics[width=1.0\columnwidth,keepaspectratio]{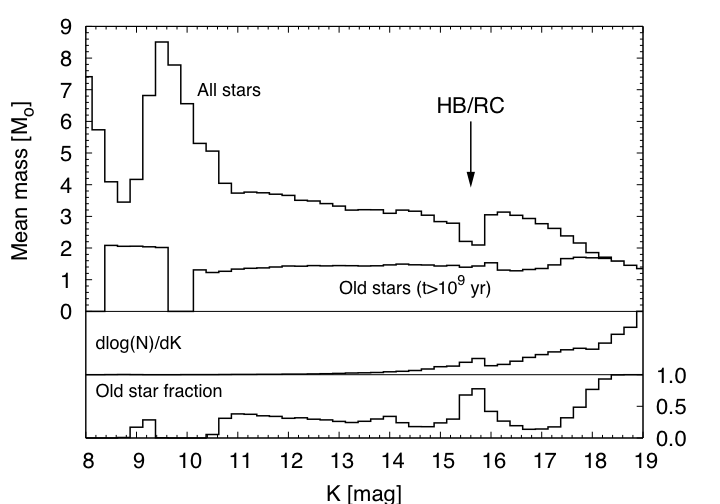}
\caption{\label{f:meanM}The predicted mean stellar mass as function of
$K$ magnitude for the total population and for the old stars based on
a population synthesis model for the central parsecs of the GC
(\citet{AlexanderSternberg1999ApJ}; \citet{Alexander2005PhR}, based on
the $Z\!=\!1.5\,Z_{\odot}$ stellar evolution tracks of
\citet{Schaller1992A&AS} and \citet{Girardi2000A&AS}).  The model
assumes continuous star formation at a constant rate over the last 10
Gyr with a Miller-Scalo initial mass function
(\citet{MillerScalo1979ApJS}). A distance to the GC of 7.62 kpc and an
extinction coefficient of $A_{K}\!=\!2.8$ mag
\citep{Eisenhauer2005ApJ} are assumed. The relative number of old
stars (defined here as stars with a lifespan longer than $10^9$ yr) is
shown in the bottom panel. The increase in the fraction of old stars
to 0.8 at $K \sim 15.5$ mag, is matched by an abrupt decrease in the
overall mean mass (top panel). This reflects the concentration of red
clump / horizontal branch (HB/RC) giants at that magnitude, as seen in
the K-band luminosity function (KLF) (middle panel).  The dotted lines
parallel to the x-axis, at y-values between 0 and 1, are merely an aid
for analysing the plot. }
\end{figure}

\begin{figure*}
\centering
\includegraphics[width=\textwidth]{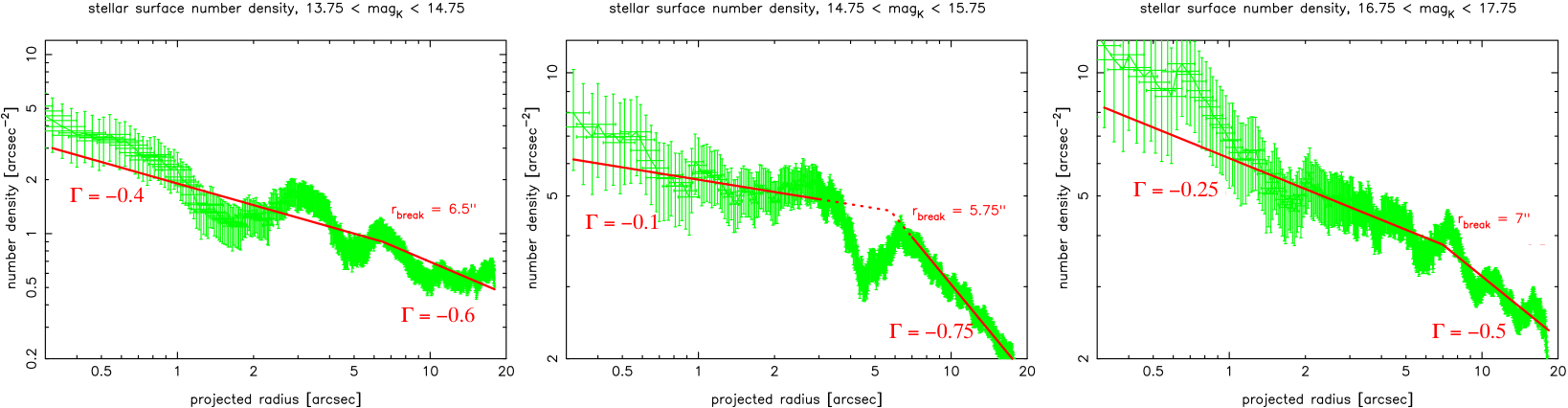}
\caption{Azimuthally averaged extinction and crowding corrected
surface density vs.\ distance from Sgr~A* for the K-magnitude ranges
$13.75-14.75$, $14.75-15.75$, and $16.75-17.75$.  Broken power-law are
fits are indicated by the red lines with the corresponding power-law
indices indicated in the panels. A 1\,$\sigma$ uncertainty of $0.05$
can be assumed for all indices, and of $0.5''$ for the break
radii. The data in the dip that can be seen in the middle panel around
$5''$ were not included in the fitting process for the intermediate
bright stars. This is indicated by the dashed line of the fitted
broken power-law in the corresponding region.
\label{Fig:densmag}
}
\end {figure*}

\begin{figure*}
\centering
\includegraphics[]{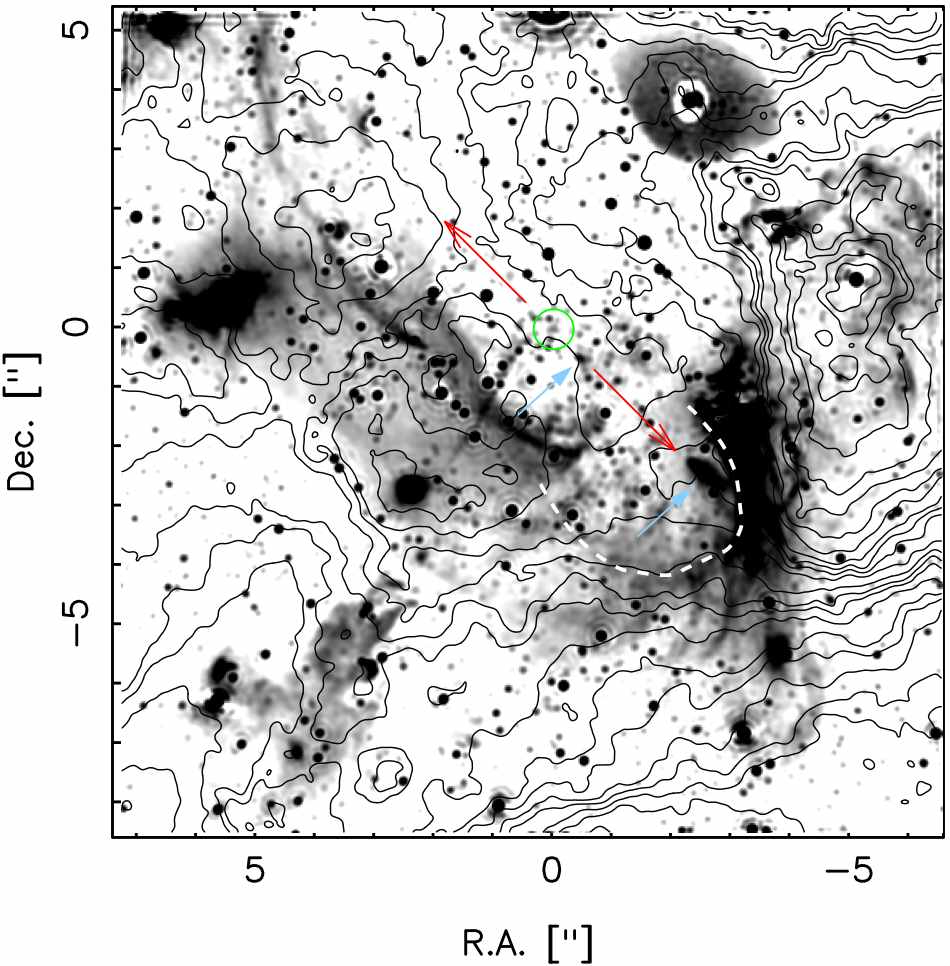}
\caption{Extinction contours (see Fig.~\ref{Fig:extNACO}) superposed
  onto a NACO L'-band image (July 2004; Lucy-Richardson deconvolved
  and restored with a Gaussian beam to a resolution of $\sim$0.1$''$;
  logarithmic scaling) of the central $10''\times10''$ of the GC. The
  mini-cavity is marked. The arrows indicate the direction of a
  possible outflow from Sgr~A*. Sgr~A* is located at the origin of the
  coordinate system and is marked by a circle. It is visible as a
  faint point source
  \citep[see][]{Genzel2003Natur,Eckart2006A&Aa,Clenet2004A&Ab,Ghez2004ApJ}.
  The (red) arrows pointing into opposing directions away from Sgr~A*
  indicate the direction of a possible outflow (see Mu\v{z}i\'{c} et
  al., in prep.). The white dashed line indicates the contours of the
  mini-cavity. The (blue) arrows (pointing up and towards the right)
  indicate cometary like features that may lie within the outflow, are
  pointed away from it and are aligned with the direction toward
  Sgr~A*. The shock-like filaments may have been created by the
  outflow (possibly expanding laterally) as well (see Mu\v{z}i\'{c} et
  al., in prep.).
\label{Fig:outflow}
}
\end {figure*}

We have used different methods to derive the structure of the GC star
cluster. These methods include analyses of star counts as well as of
the diffuse background light on seeing-limited and AO-assisted
images. All analyses have shown clearly that a single power-law is
insufficient to describe the cluster, while a broken power-law
provides an very good approximation. Table~\ref{Tab:clusterparameters}
summarises the parameters of the cluster as they were determined by
the various methods. From these independent measurements average
values of the parameters were determined with their estimated
$1\sigma$ uncertainties. These are listed in the last line of
Table~\ref{Tab:clusterparameters}.  They are a break radius
$R_{\rm break}=6.0''\pm1.0''$ ($0.22\pm0.04$\,pc), a power-law index
for the cusp of $\Gamma_{\rm cusp}=0.19\pm0.05$, and a power-law index
for the cluster outside of the break radius of $\Gamma_{\rm
cluster}=0.75\pm0.1$. These values are also in good agreement with the
values derived by the kernel method (see Fig.~\ref{Fig:kernel}).

The determined slope of the large-scale cluster is in good agreement
with what has been found already by \citet{BecklinNeugebauer1968ApJ},
and later by \citet{Catchpole1990MNRAS}, \citet{Eckart1993ApJ},
\citet{Haller1996ApJ}, \citet{Genzel2000MNRAS}, or
\citet{Launhardt2002A&A}.  Also, note that \citet{Warwick2006JPhCS}
find that a part of the diffuse X-ray radiation within $3'-12'$ around
Sgr~A* is probably caused by faint point sources with a surface
brightness that falls off as $R^{-0.87\pm0.06}$. Also,
\citet{Muno2006ApJS} find that the X-ray point sources follow a close
to isothermal power-law index in the central parsecs of the GC.

The slope of the large-scale cluster indicates therefore that it may
be close to the structure of a singular isothermal sphere. An isothermal
model for the large scale cluster would also be in good agreement with
the finding of a constant velocity dispersion of late-type stars at
projected distances outside of $0.2$\,pc \citep[see][]{Figer2003ApJ}.

As concerns the slope of the cusp, \citet{Genzel2003ApJ} derived a
value of $\Gamma=0.4\pm0.1$ for the power-law index of the projected
surface density, while we obtain a somewhat shallower cusp with
$\Gamma=0.19\pm0.05$. The shallow shape of the cusp is certainly an
important factor in explaining why it could not be found by
observations with smaller telescopes and without the aid of adaptive
optics.  Detection of the cusp is additionally hindered by the need
for a high dynamic range due to the presence of the bright, young
stars such as in the IRS~16 and IRS~13 complexes in the inner arc
seconds.  Therefore, previous publications often assumed a flat core
\citep{Eckart1993ApJ,Haller1996ApJ,Genzel1996ApJ}, with a core radius
similar to the break radius determined in this work and in
\citet{Genzel2003ApJ}.

 The kernel method (see Fig.~\ref{Fig:kernel}) gives a more detailed
impression of the structure of the inner part of the cluster, i.e.\
the cusp. The average slope derived with the kernel method agrees well
with the power-law fit described above. The kernel estimator shows
that there may be significant variations present in the slope of the
cusp, but that overall, a description by a power-law is fairly good.

The different power-law for the cusp and the improved accuracy of its
value is due to the improved data set and refined analysis in this
work. Just to mention a few improvements: use of a uniform data set,
PSF fitting with a spatially variable PSF, extinction correction,
masking of regions with extremely low completeness, no binning of the
counts, analysis of star counts as well as of the diffuse background
light.

It is interesting to note that the cluster profile based on star
counts, presented by \citet{Eckart1993ApJ}, showed a core radius of
$3.8''$. Their data were dominated by stars with a brightness $\rm
mag_{K}=14-15$, very similar to the density profile shown in the
middle panel of Fig.~\ref{Fig:densmag} that agrees remarkably well
with the results of \citet{Eckart1993ApJ}.

Although a broken power-law fit provides a satisfying description of
the structure of the GC cluster, we would like to mention that the
stellar density (Fig.~\ref{Fig:dens1017}, \ref{Fig:kernel} and
\ref{Fig:densmag}) and also the profile of the light density of the
faint stars (right panel of Fig.~\ref{Fig:light}) indicate that there
may be a separate structure present in the central arcsecond around
the black hole. This is indicated by the steeper slope of the number
and light density in this region. \citet{Eisenhauer2005ApJ} have found
that the central arcsecond appears to be dominated by B-type main
sequence stars. The crowding of stars of this peculiar type (so far
not reported in significant numbers at larger distances) supports the
idea that the region within $1''$ of Sgr~A* is dominated by a unique
cluster structure with a special stellar population. The central
arcsecond also stands out clearly in the 2D-density map in
Fig.~\ref{Fig:dens2D}. The cusp is possibly dominated by stellar black
holes \citep[see,
e.g.,][]{Morris1993ApJ,Escoude2000ApJ,Mouawad2005AN,AmaroSeoane2004MNRAS,Freitag2006ApJ,HopmanAlexander2006ApJ}.
The B-stars in the S-cluster around Sgr~A* \citep{Eisenhauer2005ApJ},
are fairly massive stars and could thus coexist with these stellar
remnants. These massive stars must be young and therefore not
dynamically relaxed. They may represent a separate entity within the
cusp and it is not yet clear how they are related to the overall
structure of the GC star cluster.

\subsection{Comparison with theory}

\subsubsection{Shape of the cluster}

A super-massive black hole is predicted to dominate the gravitational
potential within a region $r\leq r_h$, where $r_h$ is customarily
defined as the radius containing a mass in stars equal to twice the
black hole mass \citep[see][]{Merritt2006RPPh}.  As for the extent of
the cusp, the Bahcall-Wolf cusp (for a single-mass population) is
predicted to extend out to a distance of $0.1-0.2r_h$ from the black
hole \citep[see][]{Preto2004ApJ,MerrittSzell2006ApJ}.

The two-body relaxation time for a single stellar mass population,
i.e. the time for gravitational encounters to set up a locally
Maxwellian velocity distribution, is roughly
\begin{equation}
t_r = {0.34\sigma^3\over G^2m\rho\ln\Lambda}
\end{equation}
\citep{Spitzer1987book}. Within the cusp, $r\ll r_h$, the
gravitational potential is dominated by the black hole; thus
$\sigma\propto r^{-1/2}$.  Based on Figure~\ref{Fig:kernel}, the space
density in the central 0.5\,pc is reasonably well approximated by
$\rho \sim r^{-3/2}$, which makes $t_r$ approximately independent of
radius in the cusp.  Assuming solar-mass stars and setting
$\ln\Lambda\approx \ln(r_h\sigma^2/2Gm) \approx \ln\left(M_\bullet/
2m\right)\approx 14$ (Preto et al. 2004), the relaxation time in the
cusp is roughly $5\times 10^9$\,yr. It may in fact be lower by a
factor of a few because the stars in the observed brightness range are
fairly massive. Hence, the relaxation time is lower than the age of
the Milky Way.  One therefore expects to find a collisionally relaxed
distribution of orbital energies within the cusp, i.e. the
\citet{Bahcall1976ApJ} solution should hold for a single-mass stellar
population.  Core collapse, which requires a much longer time of $\sim
10^2t_r$, is not important.

The slope of $\rho(r)$ is shallower than predicted by the
(single-mass) Bahcall-Wolf form, $\rho\sim r^{-7/4}$ (Fig. 7).  One
possible reason, as discussed by \citet{MerrittSzell2006ApJ}, is that
the cusp may not have existed for a long enough time for a
steady-state energy distribution to have been reached.  Those authors
estimate that a time of nearly $10^{10}$ yr is required for the stars
around the GC black hole to attain a collisional steady state.
Depending on the time since the last event that significantly
disrupted the cusp -- e.g.  infall of an intermediate-mass black hole
-- the cusp may still be evolving toward the Bahcall-Wolf
form. Another reason for the cusp slope to be $<7/4$ is that the stars
we are probing in this work have masses in the range
$2-3$\,M$_{\odot}$ (see section~\ref{sec:masses}). If the centre of
the cluster is dominated by stellar black holes, which may have
spiralled in there due to mass segregation, then the masses of the
used tracer stars are lower than the average point mass in the nuclear
cluster, which would lead to a slope flatter than the Bahcall-Wolf
solution. Another possible explanation for the faint slope of the cusp
may be the importance of collisions because of the high density in
this region (see section~\ref{sec:enclosedmass} below).

We would like to note that we do not observe any flattening of the
cusp at the smallest radii as it would be expected because of the
onset of stellar collisions in the dense environment of the cusp
\citep[see,
e.g.,][]{Murphy1991ApJ,Alexander1999ApJ,AmaroSeoane2004MNRAS}.
\citet{Murphy1991ApJ} show that for their model 3B, which corresponds
most closely to the enclosed mass and mass densities derived in this
work, the region where the density distribution is flattened by
collisions has a radius $r\leq10^{-3}$\,pc. This corresponds to less
than $0.1''$ at the distance of the GC and can therefore not be
reasonably resolved by the present observations.

We have found a break radius -- which corresponds roughly to the
extent of the cusp -- of $6.0''\pm1.0''$ or $0.22\pm0.04$\,pc for the
GC stellar cluster (see Table~\ref{Tab:clusterparameters}). It is
important to note that this break radius is also valid for the HB/RC
population, that is, for the oldest stars in our sample that should
represent best the shape of the dynamically relaxed population. The
same break radius is also found (within the corresponding
uncertainties) for the diffuse background light, i.e., the faint,
lower-mass, unresolved stellar population (see
Figures~\ref{Fig:lightdens} and \ref{Fig:bgnaco}). From this break
radius we obtain $r_h \approx 30''-60''$ or $r_h \approx 1.1-2.2$\,pc,
if we assume that $R_{\rm break} \approx 0.1-0.2 r_{h}$.

\subsubsection{Enclosed mass \label{sec:enclosedmass}}

For estimating the enclosed mass at the GC on parsec scales, a
commonly used assumption is that the circumnuclear disk (CND) is a
rotating ring of gas, subject only to gravitation. The GC
circumnuclear disk is commonly assumed to have a radius of about
1.6\,pc and is interpreted to rotate with a circular velocity of about
110 km s$^{-1}$
\citep{Guesten1987ApJ,Jackson1993ApJ,Christopher2005ApJ}. From these
numbers, one can estimate a total enclosed mass of $\sim 4.5\times
10^6M_\odot$, of which $\sim 3.6\times 10^6M_\odot$ are due to the
black hole.  With these values $r_h$ should be roughly $4-6$ pc, or
$110-160''$, and the predicted extent of the cusp consequently a
factor of a few larger than what is actually observed.  These numbers
are of course subject to the assumption that the rotation of the CND
is a suitable tracer of the enclosed mass.

However, there are other ways to estimate the enclosed mass at large
distances. \citet{Reid2003ApJ} and Reid et al.\ (astro-ph/0612164)
present highly accurate measurements of the velocity of the maser star
IRS~9, that is located at 0.33\,pc in projection from Sgr~A*. Under
the assumption that IRS~9 is on a bound orbit, an enclosed mass of at
least $4.4\times10^{6}$\,M${_\odot}$ is required (scaling the mass
estimate of Reid et al. to the 7.6\,kpc distance of the GC assumed in
this work). This would require a much higher mass than what is derived
assuming that the CND is a suitable tracer of the enclosed mass.

\begin{figure*}
\centering
\includegraphics[width=\textwidth]{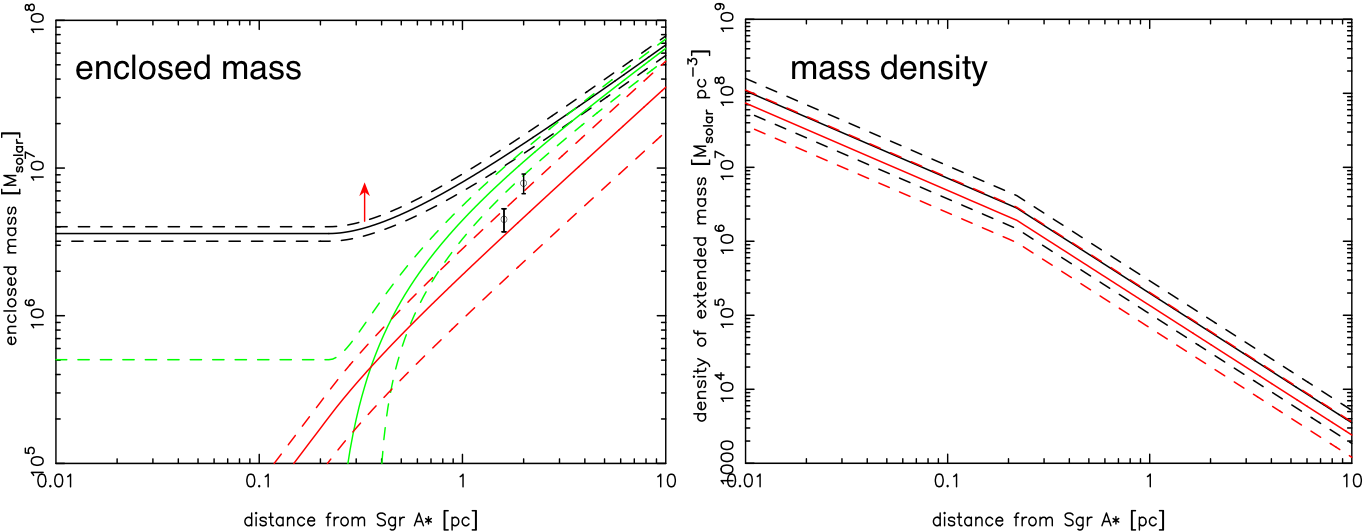}
\caption{{\bf Left panel:} Black line: Estimate of the enclosed mass
 vs.\ projected distance, derived with the Bahcall-Tremaine (BT) mass
 estimator, assuming a broken power-law structure of the stellar
 cluster and a constant line-of-sight velocity dispersion outside of
 the break radius (see text for details). The up-pointing red arrow is
 the enclosed mass estimate based on IRS~9 (Reid et al.,
 astro-ph/0612164). The circle at 1.6\,pc is the mass estimate based
 on the assupmption that the CND is a rotating ring with a rotation
 velocity of 110\,km\,s$^{-1}$ and a radius of 1.6\,pc
 \citep{Christopher2005ApJ}.  The circle at 2.0\,pc is the mass
 estimate based on the assupmption that the CND is a rotating ring
 with a rotation velocity of 130\,km\,s$^{-1}$ and a radius of 2.0\,pc
 \citep{RiekeRieke1988ApJ,Guesten1987ApJ}.  The dashed lines indicate
 the $1\sigma$ uncertainties.  Green line: enclosed mass after
 subtraction of the black hole mass, derived from the BT mass
 estimator (black). Red line: estimated mass of the visible stellar
 cluster. The dashed lines indicate the $1\sigma$ uncertainties. {\bf
 Right panel:} Density of the enclosed mass, after subtraction of the
 black hole mass (black). The red line indicates the mass density of
 the stellar cluster. The dashed lines indicate the $1\sigma$
 uncertainties.
\label{Fig:encmass}}
\end{figure*}

An alternative way of estimating the enclosed mass is using the
measured line-of-sight velocity dispersion, $\sigma_{\rm z}$, of the
old, presumably relaxed stellar population. \citet{Figer2003ApJ}
present high precision spectroscopic measurements that show that
$\sigma_{\rm z}$ is constant with a value of
$100.9\pm7.7$\,km\,s$^{-1}$, down to projected distances of at least
0.2\,pc from Sgr~A*. Similar to \citet{Figer2004ApJ}, we have used the
Bahcall-Tremaine mass estimator
\begin{equation}
M_{\rm BT}(R) = \frac{16}{\pi G N} \sum_{i}^{N} v_{\rm z}^{2} R_{i}
\end{equation}
\citep{BahcallTremaine1981ApJ} to determine the enclosed mass. Here,
$G$ is Newton's gravitational constants, $N$ is the number of stars
within a projected radius $R$ , $R_{i}$ is the projected radius of the
i-th star, and $v_{\rm z}$ its spectroscopically measured line-of-sight
velocity. Assuming isotropic stellar orbits, a constant velocity
dispersion, $\sigma_{\rm z}$, and a stellar surface density given by a
broken power-law with the parameters as listed in the last line of
Table~\ref{Tab:clusterparameters}, the enclosed mass at projected
distance $R$ is thus:
\begin{equation}
M_{\rm BT}(R) = \frac{16\sigma_{\rm z}^{2}}{\pi G} \frac{\int_{0}^{R}
(R'^{2-\Gamma}/R_{\rm break}^{-\Gamma}) dR'}{\int_{0}^{R} (R'^{1-\Gamma}/R_{\rm
break}^{-\Gamma}) dR'}
\end{equation}
The power-law index, $\Gamma$, assumes the corresponding values inside
and outside of the break radius $R_{\rm break}$. We assumed a constant
velocity dispersion outside and a Keplerian increase inside the break
radius.  The resulting mass estimate at short distances from Sgr~A*
($2.6\pm0.3\times10^{6}$\,M$_{\odot}$) was corrected to the precisely
known mass of the central black hole of
$3.6\pm0.3\times10^{6}$\,M$_{\odot}$
\citep[][]{Eisenhauer2005ApJ}. Note that the correct black hole mass
can be obtained if the break radius were located at 0.3\,pc.  The
black line in the upper panel of Fig.~\ref{Fig:encmass} shows our
estimate of the enclosed mass vs.\ projected distance.  The dashed
lines indicate the uncertainty of the measurements (it is dominated by
the uncertainty of $\sigma_{\rm z}$).  The up-pointing arrow at
$R=0.33$\,pc is the enclosed mass estimate from IRS~9 (Reid et al.,
astro-ph/0612164; their value was adapted to a GC distance of
7.6\,kpc, that is used in this work).  The error bars at $1.6$\,pc and
2.0\,pc are mass estimates based on the assumption of a circularly
rotating CND with a radius of 1.6\,pc and a rotation velocity of
110\,km\,s$^{-1}$ and a radius of 2.0\,pc and a rotation velocity of
130\,km\,s$^{-1}$, respectively
\citep[see][]{Guesten1987ApJ,Christopher2005ApJ}.  As mentioned by
\citet{RiekeRieke1988ApJ}, the uncertainties of mass estimates based
on the CND are probably large due to the presence of significant
non-circular motions within the CND.  As can be seen, the mass
estimate based on the velocity dispersion of the late-type stellar
population agrees fairly well with the lower mass estimate based on
IRS~9, but disagrees on a $>5\sigma$ level with the estimates based on
the assumption of a circularly rotating CND.  Our estimate of the
enclosed stellar mass agrees very well with the estimates by
\citet[][, see their
Figs.~8\,\&\,9]{Haller1996ApJ}. \citet{Haller1996ApJ} determined the
enclosed mass distribution for the central 300\,pc of the Milky Way.
They assumed a pseudo-isothermal structure of the stellar cluster,
with a power-law exponent of $\gamma=-1.8$, very similar to this
work. Also, they used a core radius of $7''$ for the cluster, in very
good agreement with the break radius determined by our analysis.

Similar mass estimates are obtained from stellar proper motions.
\citet{Ott2004PhDT} present an analysis of stellar proper motions
within $9''$ of Sgr~A*. They also find an approximately constant
velocity dispersion of about 100\,km\,s$^{-1}$ (one dimensional value)
outside of a projected radius of about $6''$, with a Keplerian
increase inside of this distance. They estimate the enclosed mass vs.\
projected distance from Sgr~A* with the help of various mass
estimators and always find fairly high values, in agreement with the
lower mass estimate based on IRS~9 and in agreement with the enclosed
mass estimate shown in the left panel of Fig.~\ref{Fig:encmass}. The
observation that the velocity dispersion is constant outside of $6''$
or about 0.2\,pc indicates that the gravitational field is dominated
by the point mass of the black hole only inside of this distance.

All mass estimates that are based on stellar tracers are significantly
higher than what one can derive from the assumption of a CND that is
in circular rotation. Stars are commonly accepted to be more reliable
tracers of the gravitational field than gas because they are
point-like particles and are not subject to forces from magnetic
fields, winds, or collisions. We therefore believe that the estimates
of the enclosed mass in the GC must be corrected significantly
upwards, contrary to what has been published earlier
\citep[e.g.,][]{Schoedel2003ApJ} because these earlier estimates were
based on measurements of gas velocities and an incomplete knowledge of
the velocity dispersion of the stellar cluster (a values of
$\sim$50\,km\,s$^{-1}$ was assumed in previous models).

The green line in Fig.~\ref{Fig:encmass} indicates the enclosed mass
after subtraction of the black hole mass. The red line is an estimate
of the mass of the visible stellar cluster. Here, an average mass of
3\,M$_{\odot}$ was assumed for the detected stars (see
section~\ref{sec:masses}). In order to estimate the mass of the
diffuse, unresolved component of the cluster, we assumed a
mass-to-luminosity ratio of 2\,M$_{\odot}/$L$_{\odot}/$ at 2\,$\mu$m
\citep{Haller1996ApJ}. Note that other authors \citep[e.g.\
][]{Philipp1999A&A,Kent1992ApJ} find a lower value of
1\,M$_{\odot}/$L$_{\odot}/$. This would result in a lower estimated
mass of the visible stellar cluster.  The M/L ratio was combined with
the measured surface brightness of the diffuse background light, which
was estimated to $0.007$\,Jy\,arcsec$^{2}$ at the break radius. The
systematic uncertainty of the background light density is about 30\%
as can be estimated by a comparison of Figs.~\ref{Fig:lightdens} and
\ref{Fig:bgnaco}. The main sources of the uncertainty are the choice
of the calibration sources and the background subtraction in the ISAAC
data. To correct for the nuclear bulge, a surface brightness of
$0.0006$\,Jy\,arcsec$^{2}$ was subtracted \citep[value at $\sim$10\,pc
taken from][]{Philipp1999A&A}.

The black line in the right panel of Fig.~\ref{Fig:encmass} shows the
volume density of the estimated enclosed mass:
\begin{equation}
2.8\pm1.3\times 10^{6} {\rm M_{\odot}\,pc^{-3}} \times \left(\frac{r}{0.22\,{\rm
  pc}}\right)^{-\gamma},
\end{equation}
where $r$ is the distance from Sgr~A*, $\gamma=1.2$ the power-law
index inside $0.22$\,pc and $\gamma=1.75$ the power-law index outside
of $0.22$\,pc.  The resulting densities are similar to the values
given in \citet{Genzel2003ApJ}. The red line shows the estimated
density of the visible stellar cluster. The dashed lines indicate the
$1\sigma$ uncertainties. The data show that the estimated mass density
is possibly a factor of roughly $2$ higher than the density of the
visible stellar cluster. Of course, the uncertainties are large. They
can be imporved by more precise measurements of the diffuse surface
brightness, the velocity dispersion, and of the central black hole
mass.  The inferred mass densities are in good agreement with the
extent of the cusp derived in this work. The left panel of
Fig.~\ref{Fig:encmass} shows that the cluster contains twice the mass
of the black hole within a projected radius of $<$2\,pc. This is in
good agreement with the $r_h=1.1-2.2$ predicted from the measured
extension of the cusp ($0.22\pm0.04$\,pc). Also, the fairly flat
power-law of the cusp $\Gamma=1.2\pm0.05$ may be related to a high
central density because flat cusps are created when collisions become
important \citep{Murphy1991ApJ}.

There are certain caveats to keep in mind, i.e. basic assumptions that
we used for our mass estimate: (a) The assumption of a constant
velocity dispersion outside of 0.22\,pc.  This assumption seems,
however, well justified because our estimation of the enclosed mass
agrees well with the data by \citet{Haller1996ApJ}, who used velocity
dispersion measurements out to 300\,pc \citep[see,
however,][]{Launhardt2002A&A}. Also, the basic conclusion, that there
is possibly a significant amount of non-stellar matter present stays
valid if we limit the analysis only to distances within 1\,pc, based
on the velocity dispersion measurements by \citet{Figer2003ApJ} and
\citet{Genzel2000MNRAS}. (b) Assumption of the given structure of the
cluster. As for the visible cluster, a power-law index of $1.8$ is,
however, well established \citep[see discussion in this work and,
e.g.,][]{BecklinNeugebauer1968ApJ,Catchpole1990MNRAS,Eckart1993ApJ,Genzel2003ApJ}.
Variations of the cluster slope within the uncertainties will not have
any significant influence on our estimate.  The extent and power-law
index of the cusp are only of minor importance, due to its small
extent and the small total mass enclosed in this volume. (c) Validity
of the BT mass estimator. The use of the BT mass estimator seems
justified because it provides the correct mass for the central black
hole (within a factor of 30\%) and it agrees well with other mass
estimators \citep[see, e.g., tests by][]{Ott2004PhDT}. (d) Assumption
of a given M$/$L ratio for the stellar cluster. It seems, however,
well established and we have used the more conservative one of the
values,i.e. the one that results in a larger stellar mass, established
by observations \citep[see][]{Philipp1999A&A,Kent1992ApJ}.

What can the enclosed extended dark mass be composed of? Primary
candidates are dark matter and stellar remnants. As for dark matter
\citet{Bergstrom2006PhRvD} inferred upper limits on the density of
dark matter using the density profile of cold dark matter of
\citet{Navarro1996ApJ}. The resulting density of
66\,M$_{\odot}$\,pc$^{-3}$ at a distance of 1\,pc from Sgr~A* is by
several orders of magnitude too small in order to have a measurable
influence.

Due to their high mass, stellar-mass black holes sink toward the GC
via dynamical friction.  Several authors have estimated the amount of
mass in form of stellar-mass BHs in the central parsec
\citep[e.g.,][]{Morris1993ApJ,Escoude2000ApJ}. \citet{Escoude2000ApJ}
find that about $25,000$ stellar-mass BHs may be present in the
central parsec, corresponding to roughly
$2.5\times10^{5}$\,M$_{\odot}$. Similarly low numbers are found by
\citet{Freitag2006ApJ} and \citet{HopmanAlexander2006ApJ}. A much
higher value is derived by \citet{Morris1993ApJ}, who calculate that
of the order $10^{6}$\,M$_{\odot}$ may be present in form of
stellar-mass black holes in the inner tenths of a parsec. These higher
estimates are probably due to the stellar mass function assumed by
\citet{Morris1993ApJ}, which contain a larger fraction of stellar
black hole progenitors.

The publications cited above considered mainly inflow of heavy stellar
remnants from the inner bulge over the life time of the Galaxy,
starting from an initially present stellar population with a given
IMF. They did not consider possibly continually on-going star
formation in the GC. \citet{Figer2004ApJ} infer a star formation rate
of about $4\times10{-7}$ \,M$_{\odot}$\,yr$^{-1}$\,pc$^{-3}$ in the
inner 50\,pc of the GC. Under the assumption that this star formation
rate is a typical value and has been largely constant over the life
time of the Milky Way, this corresponds to a total mass of stars of
$1.7\times10^{7}$\,M$_{\odot}$ formed in the inner 10\,pc over
$10^{10}$\,yr. Of course, a large fraction of the mass bound in stars
will be returned to the ISM. The lighter stellar remnants may be
pushed outward by mass segregation. However, if we assume that stars
heavier than 30\,M$_{\odot}$ form stellar mass black holes of
7\,M$_{\odot}$ \citep[see][]{Escoude2000ApJ}, then
$\sim$$1.2\times10^{6}$\,M$_{\odot}$ will be locked in stellar-mass
black holes. Here, we have assumed an IMF of the form $N(m)\propto
m^{-1.3}$ and an upper mass cut-off of 120\,M$_{\odot}$ (choosing
80\,M$_{\odot}$ will give a very similar result).

We also have to consider the star formation that takes place in the
inner 0.5\,pc.  \citet{Paumard2006ApJ} estimate a mass of
$\sim$$10^{4}$\,M$_{\odot}$ for the young stars that formed in the
most recent star burst event(s).  There is evidence for another star
formation event $10^{8}$\,yr ago \citep[see][]{Krabbe1995ApJ}. If
these values are typical, then we obtain a star formation rate of
$10^{-4}$\,M$_{\odot}$\,yr$^{-1}$ for the central half parsec. With
the same assumptions used above, this leads to roughly
$7\times10^{5}$\,M$_{\odot}$ locked in stellar-mass black holes.
Since the IMF in the central parsec appears to be top-heavy
\citep{Nayakshin2005MNRAS,Paumard2006ApJ}, the efficiency of producing
black holes may be considerably higher in the GC. Note the dynamical
upper limit on the number of stellar-mass black hole that can be
packed into the central 0.5 pc for 10 Gyr while avoiding kicking each
other into the MBH \citep[the "drain limit"][]{AlexanderLivio2004ApJ}
is not much larger, $\sim 2\times10^{5}$.

The considerations above are certainly very rough estimates, but show
that it is not unreasonable to assume that a large number of dark
stellar remnants may be present in the central parsecs of the GC.  In
fact, \citet{Muno2005ApJ} have found evidence that X-ray transients
are overabundant by a factor of $\gtrsim20$ per unit stellar mass in
the central parsec. If a significant fraction of these transients is
related to stellar mass black holes, this may explain the excess in
non-stellar mass in the GC cluster.  There are also contradictory
studies. \citet{Deegan2006JPhCS} conclude from considerations
concerning the diffuse X-ray emission from the central parsec that not
more than $20,000$ stellar-mass BHs can be present in this
region. Clearly, the issue needs further intensive study. Here, we
mainly want to make the point that stellar dynamics shows that the
amount of mass in the extended stellar cluster appears to be much
larger than what could be explained by stars alone. Stellar remnants
may be a possibility to explain this discrepancy. A large number of
stellar remnants would have important implications on physical
mechanisms such as gravitational lensing of background stars, and
destruction and scattering of stars in the central parsec \citep[see,
e.g.,][]{Morris1993ApJ,Alexander2001ApJ,Alexander2005PhR}.

\section{Summary}

In this work, we have presented seeing limited and AO assisted
high-resolution NIR imaging observations of the stellar cluster in the
central parsec of the GC. The data were used to extract information on
the stellar number density and on the diffuse background light density
in the GC. Both star counts and background light density were
corrected for crowding effects and interstellar extinction in the GC.
A detailed map of the variation of the interstellar extinction in the
central parsec of the Milky Way was presented. We compared our
analysis to theoretical expectations. Our results are summarised in
the following points:

\begin{enumerate} 
\item The overall structure of the GC nuclear cluster can be described
  well by a broken power-law as found already by
  \citet{Genzel2003ApJ}. A single-power-law is clearly insufficient to
  describe the cluster structure.
\item From the different data sets and methods (background light,
  number counts), we have derived various estimates of the inner and
  outer power-law indices and of the break radius. The different
  estimates agree well within their uncertainties. From their
  comparison, we derive a power-law index of $\Gamma=0.19\pm0.05$ for
  the cusp, a power-law index of $\Gamma=0.75\pm0.1$ for the outer
  part of the cluster, and a break radius of $R_{\rm break} =
  6.0\pm1.0''$ or $0.22\pm0.04$\,pc (the power-law indices refer to
  the measured {\it projected} density). The cusp appears flatter than
  what is predicted by most theoretical work. This may be related to a
  very high density of the stellar cluster.
\item The best available measurements of the velocity dispersion of
  the cluster from the literature are used in order to derive the
  enclosed mass/mass density vs.\ distance from Sgr~A*. We obtain
  significantly higher mass estimates for the central parsec than
  previously considered. The inferred density of the extended mass at
  the break radius is $2.8\pm1.3\times 10^{6} {\rm
  M_{\odot}\,pc^{-3}}$. The inferred high mass of the cluster agrees
  very well with the extent of the cusp, assuming that the cusp
  extends to $0.1-0.2 r_{h}$, where $r_{h}$ is the radius around
  Sgr~A* within which the cluster contains twice the mass of the
  supermassive black hole.  We find that the mass of the stars in the
  cluster may only represent about 50\% percent of the actually
  present mass in the cluster within the central pc. The work of other
  authors and some rough estimates in this work show that stellar-mass
  black holes may contribute a large part of the dark mass.
\item The examination of the stellar surface density for different
  magnitude bins reveals a trend in the spatial distribution that
  could be the result of mass segregation. The horizontal branch
  stars, with K-magnitudes around 15.25, are in their majority
  horizontal branch/red clump stars that are expected to trace the
  old. relaxed stellar population. They have the lowest mean mass of
  all observed stars in our images and, consistent with theoretical
  expectations, the flattest power-law index inside the cusp. We also
  find that there exists either an under-density of HB/RC stars around
  $R=5''$, or, alternatively, an over-density of similarly bright
  stars at $R=3''$ and $R=7''$.
\item Several density clumps appear to be present in the cluster,
  especially at projected distances of $3''$ and $7''$ from
  Sgr~A*. Their nature is unclear, but they may be related to the
  recent star formation events.
\item A detailed map of the variation of interstellar extinction in
  the central $\sim0.5$\,pc of the Milky Way is presented. Extinction
  is highly variable and has a general minimum centred on Sgr~A* and
  in a strip running NE-SW across Sgr~A*. The extinction map in
  combination with possible shock structures in an L'-band map may
  provide some support to the assumption of an (aligned) outflow from the
  central arc seconds. 
\end{enumerate}

\begin{acknowledgements}
      Part of this work was supported by the German \emph{Deut\-sche
      For\-schungs\-ge\-mein\-schaft, DFG\/} project number SFB 494
      and by the DFG Schwerpunktprogramm SPP1177. TA was supported by
      ISF grant 928/06, Minerva grant 8563, and a New Faculty grant
      by Sir H.\ Djangoly, CBE, of London, UK. DM was supported by
      grants AST-0437519 from the NSF and NNG04GJ48G from NASA.
\end{acknowledgements}

\bibliography{gc}

\newpage

%

\begin{table}
\caption{Details of the ISAAC/VLT observations used for this work. DIT
  is the detector integration time. NDIT is the number of integrations
  that were averaged online by the read-out electronics. N is the
  number of exposures that were taken. The total integration time
  amounts to N$\times$NDIT$\times$DIT. FWHM is the approximate full
  width at half maximum of the PSF due to seeing. The pixel scale was
  $0.147''$ per pixel.
\label{Tab:ISAAC}}
\centering
\begin{tabular}{lllllll}
\hline\hline
Date & $\lambda_{central}$\,[$\mu$m] &  $\Delta\lambda$\,[$\mu$m] & N
& NDIT & DIT [sec] & FWHM\\
\hline
02 July 1999 &  1.25 & 0.29 & 6 & 12 & 24.0 & $0.4''$ \\
02 July 1999 &  2.09 & 0.02 & 6 & 12 & 24.0 & $0.4''$ \\
27 July 2005 &  1.65 & 0.30 & 8 & 5 & 1.8 & $0.9''$ \\
\hline
\end{tabular}
\end{table}

\begin{table}
\caption{Details of the NACO/VLT observations used for this work.DIT
  is the detector integration time. NDIT is the number of integrations
  that were averaged online by the read-out electronics. N is the
  number of exposures that were taken. The total integration time
  amounts to N$\times$NDIT$\times$DIT. The pixel scale of all
  observations was $0.027''$ per pixel.
\label{Tab:Data}}
\centering
\begin{tabular}{lllllll}
\hline\hline
Date & $\lambda_{central}$\,[$\mu$m] &  $\Delta\lambda$\,[$\mu$m] & N & NDIT & DIT [sec]\\
\hline
12 June 2004  & 2.06 & 0.06  & 96   & 1  & 30\\
12 June 2004  & 2.24 & 0.06  & 99   & 1  & 30\\
9 July 2004  & 2.00 & 0.06  & 8   & 4  & 36\\
9 July 2004  & 2.24 & 0.06  & 8   & 4  & 36\\
9 July 2004   & 2.27 & 0.06  & 8   & 4  & 36\\
9 July 2004   & 2.30 & 0.06  & 8   & 4  & 36\\
\hline
\end{tabular}
\end{table}

\begin{table}
\caption{Completeness of stars detected in the ISAAC $2.09$\,$\mu$m
image. $R$ is the projected distance from Sgr~A*.
\label{Tab:ISAAC_completeness}}
\centering
\begin{tabular}{lllllll}
\hline\hline
mag$_{\rm K}$ & $R<60''$ & $15''<R<60''$ & $R<15''$ \\
\hline
12 &  99\% & 99\% & 98\% \\ 
13 &  99\% & 99\% & 93\% \\ 
14 &  95\% & 97\% & 80\% \\ 
15 &  87\% & 89\% & 53\% \\
16 &  63\% & 66\% & 16\% \\ 
17 &  22\% & 23\% &  1\% \\ 
\hline
\end{tabular}
\end{table}

\begin{table}
\caption{Completeness of stars detected in the NACO $2.27+2.30$\,$\mu$m
image. $R$ is the projected distance from Sgr~A*.
\label{Tab:NACO_completeness}}
\centering
\begin{tabular}{lll}
\hline\hline
mag$_{\rm K}$ & $R<8''$ & $R>8''$ \\
\hline
14 &  97.1\% & 98.7\% \\ 
15 &  93.5\% & 96.9\% \\
16 &  84.4\% & 91.6\% \\
17 &  65.1\% & 77.8\% \\
\hline
\end{tabular}
\end{table}

\begin{table}
\caption{Break radius and power-law indices for the central stellar
cluster.
\label{Tab:clusterparameters}}
\centering
\renewcommand{\footnoterule}{}
\begin{tabular}{llll}
\hline\hline
method & R$_{\rm break}$ & $\Gamma_{\rm cusp}$ & $\Gamma_{\rm cluster}$ \\
\hline
counts$_{\rm ISAAC}^{\mathrm{a}}$ & - & - & $0.68\pm0.02$ \\
background$_{\rm ISAAC}^{\mathrm{b}}$ & $6.5\pm1.1$  & $0.2\pm0.11$ & $0.95\pm0.11$ \\
counts$_{\rm NACO}^{\mathrm{c}}$ & $6.75\pm0.5$ & $0.30\pm0.05$ &  $0.55\pm0.05$ \\
background$_{\rm NACO}^{\mathrm{d}}$ & $4.0\pm0.8$ & $0.15\pm0.07$ &  $0.85\pm0.1$ \\
counts$_{\rm NACO,HB}^{\mathrm{e}}$ & $5.75\pm0.5$ & $0.10\pm0.05$ &  $0.75\pm0.05$ \\
\hline
 & $6.0\pm1.0$ & $0.19\pm0.05$ & $0.75\pm$0.10 \\
\hline
\end{tabular}
\begin{list}{}{}
\item[$^{\mathrm{a}}$] star counts on ISAAC image (see Fig.~\ref{Fig:ISAAC_density} and
  section~\ref{sec:ISAAC_counts})
\item[$^{\mathrm{b}}$] diffuse background on ISAAC image (see Fig.~\ref{Fig:background} and
  section~\ref{sec:bglight}); uncertainties based on formal fit error
  and tests with artificial background
\item[$^{\mathrm{c}}$]  star counts on NACO image (see Fig.~\ref{Fig:dens1017} and
  section~\ref{sec:density})
\item[$^{\mathrm{d}}$] diffuse background on NACO image (see Fig.~\ref{Fig:bgnaco} and
  section~\ref{sec:countsvsdensity})
\item[$^{\mathrm{e}}$]  star counts on NACO image, brightness range of
  HB/RC stars (see Fig.~\ref{Fig:densmag} and
  section~\ref{sec:masses})
\end{list}
\end{table}

\newpage



\clearpage

\end{document}